\pdfoutput=1

\documentclass[11pt]{article}
\usepackage{amsmath}             
\usepackage{geometry,bbm,nicefrac,comment,fontawesome,cases}
\usepackage[ruled,linesnumbered]{algorithm2e}
\usepackage{array}
\usepackage{setspace} 
\usepackage{color}
\usepackage[final]{graphics}
\usepackage[draft]{graphicx}          	 
\usepackage{bm}                 
\usepackage{amssymb}
\usepackage{amsfonts}             
\usepackage{verbatim}           
\usepackage{amsthm}             
\usepackage{enumerate}
\usepackage{enumitem}
\usepackage{mathtools} 
\usepackage{relsize}
\usepackage{hyperref}
\usepackage{cleveref}
\usepackage{authblk}   
\usepackage{verbatim}   
\usepackage{subcaption}

\usepackage[title]{appendix}
\numberwithin{equation}{section}	
\usepackage{multirow}
\usepackage{booktabs}
\usepackage{caption}
\usepackage{diagbox}
\theoremstyle{plain}             

\newtheorem{remark}{Remark}

\providecommand{\keywords}[1]
{
  \small	
  \textbf{\textit{Keywords:}} #1
}

\def\d{{\rm d}}
\def\B{{\mathbf{B}}}
\def\X{{\mathbf{X}}}
\def\Cov{{\rm Cov}}
\def\Var{{\rm Var}}
\def\dsp{\displaystyle}
\def\FG#1#2#3#4{
{}_2F_1\left(
\begin{array}{c}
\begin{array}{c}\hskip-10pt#1,\,#2\end{array}\\
\begin{array}{c}\hskip-10pt #3\end{array}
\end{array}
\hskip-8pt;\,#4
\right)}

\newcommand{\sm}{\mbox{(semi-)}}

\begin{document}

\title{Solution of integrals with fractional Brownian motion for different Hurst indices}
\author[a]{Fei Gao}
\author[b]{Shuaiqiang Liu}
\author[c]{Cornelis W. Oosterlee \thanks{Corresponding author : c.w.oosterlee@uu.nl}}
\author[b]{Nico M. Temme}

\affil[a]{School of Mathematics and Statistics, Xi'an Jiaotong University}
\affil[b]{Research Group of Scientific Computing, Centrum Wiskunde \& Informatica}
\affil[c]{Mathematical Institute, Utrecht University}
\renewcommand*{\Affilfont}{\small\it}
\renewcommand\Authands{, } 
\date{} 

\maketitle
\begin{abstract}
In this paper, we will  evaluate integrals that define the conditional expectation, variance and characteristic function of stochastic processes with respect to fractional Brownian motion (fBm) for all relevant Hurst indices, i.e. $H \in (0,1)$.  The fractional Ornstein-Uhlenbeck (fOU) process, for example, gives rise to highly nontrivial integration formulas that need careful analysis when considering the whole range of Hurst indices. We will show that the classical technique of analytic continuation, from complex analysis, provides a way of extending the domain of validity of an integral, from $H\in(1/2,1)$, to the larger domain, $H\in(0,1)$. 
Numerical experiments for different Hurst indices, confirm the robustness and efficiency of the integral formulations presented here. Moreover, we provide accurate and highly efficient financial option pricing results for processes that are related to the fOU process, with the help of  Fourier cosine expansions.
\end{abstract} \hspace{10pt}

\keywords{Fractional Brownian motion, Fractional Ornstein-Uhlenbeck process, Conditional density function, Analytic continuation, Numerical approximation, COS method option pricing}

\section{Introduction}
In this paper, we will  evaluate the integrals,  that define the conditional expectation and variance of stochastic processes that are based on fractional Brownian motion (fBm). For this, we first define fBm, to which a Hurst index is associated, as a generalization of standard Brownian motion (sBm). 

Let $(\Omega,\Sigma,\mathbb{P})$ be a complete probability space, $\B^H=\{B_t^H,\,t\geq 0 \}$ be an fBm with Hurst index $H\in(0,1)$ defined on this probability space, and 
$\mathcal{F}_t$ be its natural filtration.

Fractional Brownian motion,
$\B^H$, is then uniquely characterized by the following properties \cite{norros1994storage}: 
\begin{enumerate}[label=\arabic*),leftmargin=3em]
	\item $\B^H$ has stationary increments, which means $B^H_t - B^H_s \sim B^H_{t-s}$, for $ 0 \leq s \leq t$; 
	\item $B^H_0=0$ and $\mathbb{E}[B^H_t]=0$, for $t\geq0$;
	\item $\mathbb{E}[B^H_t]^2=t^{2H}$, for $t\geq0$;
	\item $B^H_t$ has a Gaussian distribution, for $t>0$.
\end{enumerate}
From the first three properties, it follows that the covariance function is given by
\begin{equation}
\Cov\,(B^H_t,B^H_s)=\frac{1}{2}\,(|t|^{2H}+|s|^{2H}-|t-s|^{2H}), \;\; s \le t.
\label{cov}
\end{equation}
The special case of \Cref{cov} is sBm ($H=1/2$), with covariance function $\Cov\,(B^{\frac{1}{2}}_t,B^{\frac{1}{2}}_s)$ $={\rm min}\,(t,s)$. 
 Particularly for fBm, and different from sBm, the increments are not independent and fBm is neither a Markov process nor a semi-martingale, for $H\neq 1/2$. As a consequence, fBm enables modeling of stochastic processes that are self-similar or exhibit long-range dependence in time. Properties of fBm have been presented, for example, in \cite{nualart2006fractional} and \cite{biagini2008stochastic}.

Fractional Brownian motion forms the basis for a broad class of stochastic processes, like the fractional Ornstein-Uhlenbeck (fOU), the fractional Cox-Ingersoll-Ross (fCIR) and the geometric fOU (GfOU) processes. Such processes have played an increasingly important role in diverse application fields, like hydrology
\cite{molz1997fractional,lu2003efficient}, telecommunication \cite{norros1995use,pashko2017simulation}, weather forecast \cite{tsonis1999long,rivero2016new} or epidemic disease (such as the Corona virus) modeling \cite{akinlar2020solutions}. FBm has also been used in finance for the modeling of commodities, weather (now considered as an asset) or currencies in financial products, that are based on these assets, in \cite{power2010long,brody2002dynamical, benth2003arbitrage, benth2012modeling,xiao2010pricing}, for bond markets in \cite{fink2013conditional} and recently for rough volatility in \cite{gatheral2018volatility,livieri2018rough}.
In these application areas, we require the computation of conditional expectations, which relate, next to the average behaviour under uncertainty, also to the valuation of options on these non-tradable assets in finance.

For Markov processes, the accurate computation of conditional expectations and variances has been given ample attention. In the diverse literature, we encounter, for such processes, approaches based on numerical integration, Fourier or Laplace inversion, Monte Carlo simulation and recently neural network approximations. For non-Markov processes, to which fBm belongs, the literature is more scarce but some significant books and papers have been presented in the last twenty years, like~\cite{nualart2006fractional,biagini2008stochastic,benth2003arbitrage,fink2013conditional,fink2011fractional,dieker2004simulation}. The conditional characteristic function for the fOU process has been derived in \cite{fink2013conditional} and was evaluated in \cite{fink2013conditional,wang2021modeling}, where involved integration formulas had to be calculated numerically. These integration formulas are nontrivial when considering all Hurst indices $H\in(0,1)$, due to occurring singularities in the integrand. Particularly, for the range $H \in(0,1/2)$, we encounter approximation formulas in the literature that are not easily understood from a numerical point-of-view.

In this paper, we therefore reconsider the expectation and variance related integration formulas for all relevant Hurst indices, and aim to approximate them numerically
on the basis of accurate (classical) approximations and numerical techniques.
The technique of analytic continuation~\cite{Titchmarsh:1958:TTF, Roy:2010:AAM} is used in this paper to handle the integration for certain Hurst index values. Analytic continuation provides a way of extending the domain over which a complex function is defined.
Here, it is used to extend the domain of validity of an integral which is defined on an interval with respect to the Hurst index, from $H\in(1/2,1)$, to the larger domain, $H\in(0,1)$. Some examples and pointers for  details on analytic continuation are presented in \ref{appendix.analytic continuation}.
Here, we calculate the conditional expectation, variance and the conditional characteristic function of fBm and fBm driven stochastic differential equations (SDEs). The accuracy of our numerical results is verified by results obtained by Monte Carlo simulation.

As an application in finance, we discuss the option pricing problem where the asset price is modeled as a fractional stochastic process. Clearly, fBm is not a semi-martingale, but some assets, like those based on the weather or volatility, are non-tradable products so that the well-known financial derivative pricing theory, based on the no-arbitrage and market completeness, cannot be used anyway. Valuation of such a derivative is then  based on the expected discounted value approach under the real world probability measure. With the characteristic function for the fOU process available, we can compute the conditional expectation, by means of the COS method, a Fourier-based numerical integration technique, see \cite{fang2009novel}, \cite{fang2009pricing}. The  COS method is based on an approximation of the appearing (conditional) probability density function by a Fourier cosine series expansion. The characteristic function of some of the fractional processes considered here is either not available or not easy to derive, however,  by means of a transformation of variables, the COS method can still be employed.

The remainder of this paper is organized as follows. \Cref{sec.2} briefly introduces fBm and the related stochastic processes. In \Cref{sec.3}, we analyze the conditional variance of the fOU process in detail and present a convenient numerical method to compute the appearing integrals. Option pricing with the COS method is described in \Cref{sec.option pricing}. \Cref{sec.5} presents numerical results including conditional distributions and option pricing. \Cref{sec.6} gives our conclusions of the paper.

\section{Fractional Brownian motion and related  processes}
\label{sec.2}
In this section, we discuss some well-known properties of fBm and also several stochastic processes driven by fBm.

\subsection{Fractional Brownian motion}
\begin{figure}[!htb]
	\centering
	\includegraphics[height=6cm]{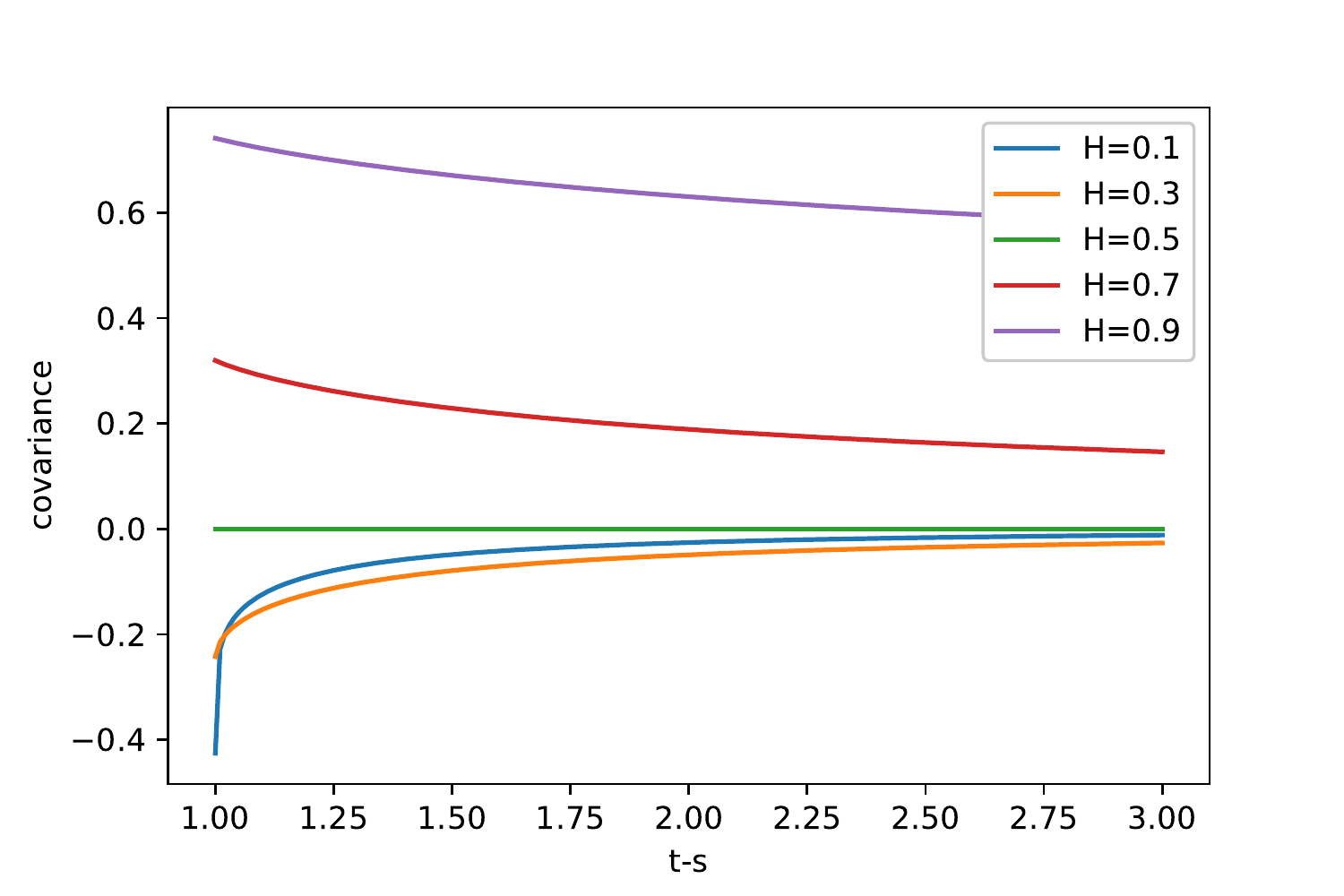}
	\caption{Covariance of fBm increments for various $H$.}
	\label{fig.cov}
\end{figure}

Fractional Brownian motion $\B^H$ has the following three important properties.
\begin{enumerate}[label=\arabic*),leftmargin=3em]
	\item \textit{Self-similarity}. $\B^H$ is self-similar with Hurst index $H$, which means $B^H_{pt}$ and $p^{H}B^H_t$ have the same finite-dimensional distributions, for all $p>0$.
	
	\item \textit{Regularity}. Trajectories of $\B^H$  are locally Hölder continuous of any order strictly less than $H$, which means for every $\varepsilon > 0$, there exists a (random) constant $C$ such that $|B^H_t-B^H_s|\leq C\,|t-s|^{H-\varepsilon}$.
	
	\item \textit{Long-range dependence}. 
	 The autocovariance function of the increment process, $\mathbf{Y}=\{Y_t,\, t\geq 0 \}$, defined by 
	$Y_t := B^H_{t+1}-B^H_t$, reads,
	\begin{equation}
	\begin{aligned}
	\Cov\,(Y_t,Y_s) & = \mathbb{E}[(B_{t+1}^H-B_t^H)(B_{s+1}^H-B_s^H)] \\& =\frac{1}{2}\,[(t-s+1)^{2H}-2(t-s)^{2H}+(t-s-1)^{2H}],
	\end{aligned}
	\label{covariance}
	\end{equation}
	where $0 \leq s \leq t$. The function $\Cov\,(Y_t,Y_s)$ is presented, for various $H$, in \Cref{fig.cov}. $\Cov\,(Y_t,Y_s)=0$ for any $t$ and $s$, when $H=1/2$, since sBm has independent increments, whereas increments of fBm are not independent. The increments are positively correlated when $H>1/2$ and negatively correlated when $H<1/2$. Moreover, when $H>1/2$, $\Cov\,(Y_t,Y_s)$ decays very slowly with increasing time increment $(t-s)$, and therefore fBm exhibits long-range dependence when $H>1/2$.
\end{enumerate}
	
\subsection{Fractional Ornstein-Uhlenbeck process}
\label{sec.foup}
The fractional Ornstein-Uhlenbeck (fOU) process, $\X=\{X_t,\, t\geq 0 \}$, is based on fBm and is defined by the following SDE:
\begin{equation}
\d X_t = \lambda(\mu-X_t)\,\d t + \sigma \, \d B^H_t,\quad t\in[0,T],
\label{foup}
\end{equation}
with an fBm $\B^H$ and given initial value $X_0$. The parameter $\mu$ represents the long-term mean value of the process, $\lambda$ the speed of mean reversion, and $\sigma$  the volatility. It has a unique path-wise solution as 
\begin{equation}
X_t = X_0 \, e^{-\lambda t}+\mu (1-e^{-\lambda t})+\sigma \int_{0}^{t} e^{- \lambda  (t-s)} \, \d B^H_s.
\end{equation}
For $H=1/2$, using It$\hat{\rm o}$'s isometry property, the analytical solution of the Ornstein-Uhlenbeck process driven by standard Brownian motion reads:
\begin{equation}
X_t \stackrel{d}{=} X_0 \, e^{-\lambda t}+\mu (1-e^{-\lambda t})+\sigma \sqrt{\frac{1-e^{-2\lambda t}}{2\lambda}}W,
\end{equation}
where $W$ is a sample from the standard normal distribution $\mathcal{N}(0,1)$.

Using $ \kappa :=H-1/2$, and based on the derivation of conditional characteristic function, it is proved in \cite{fink2013conditional} that under the filtration $\mathcal{F}_s \, (s \geq 0)$, $X_t \, (t \geq s)$ is normally distributed, with
\begin{equation}
\mathbb{E}[X_t|\mathcal{F}_s] = X_s \, e^{-\lambda(t-s)}+\mu(1-e^{-\lambda(t-s)})+\int_0^s \Psi_c(s,t,v) \, \d B_v^H,
\label{fou.mean}
\end{equation}
\begin{equation}
\Var[X_t|\mathcal{F}_s] = \Vert c(r) \textbf{1}_{[s,t]}(r) \Vert _{\kappa,T}^2 - \Vert \Psi_c(s,t,v) \textbf{1}_{[0,s]}(v) \Vert _{\kappa,T}^2,
\label{fou.var}
\end{equation}
where $c(r)=\sigma e^{-\lambda(t-r)}$; and, for $v \in (0,s)$,
\begin{equation}
\Psi_c(s,t,v)=\frac{\sin(\pi \kappa)}{\pi}v^{-\kappa}(s-v)^{-\kappa}\int_s^t \frac{r^\kappa(r-s)^\kappa}{r-v}c(r)\, \d r,
\end{equation}	
while for $v \in\{0,s\}$, $\Psi_c(s,t,v)=0$. For $\kappa \neq 0$, the norm of a function $f$ is defined by the fractional Riemann-Liouville integral
\begin{equation}
\Vert f \Vert _{\kappa,T}^2 = A_{\kappa} \int_0^T z^{-2\kappa}\left(\int_z^T \frac{r^\kappa f(r)}{(r-z)^{1-\kappa}}\,\d r\right)^2\,\d z,
\label{eq.norm}
\end{equation}
with 
\begin{equation}
A_{\kappa}=\frac{\pi \kappa(2\kappa+1)}{\Gamma (1-2\kappa)\sin(\pi \kappa) \Gamma^2(\kappa)},
\label{A_kappa}
\end{equation} 
where $\Gamma(\cdot)$ is the gamma function.
For $\kappa = 0$, $\Vert f \Vert _{0,T}^2$ is set as the $L_2$ norm, i.e.,
\begin{equation}
\Vert f \Vert _{0,T}^2 = \int_0^T f^2(r)\, \d r.
\end{equation}

The accurate numerical evaluation of \Cref{fou.mean} and \eqref{fou.var} is a focus of the present paper. These equations also form the basis for the conditional characteristic function.

\begin{remark}[Conditional distribution of fBm]
The fBm can be considered as a special case of the fOU process by taking $\lambda=0$ and $\sigma=1$ in \Cref{foup}. Then, the conditional distribution of the fBm is also a normal distribution, with
\begin{equation}
\mathbb{E}[B^H_t|\mathcal{F}_s] = B^H_s+\int_0^s \Psi(s,t,v) \, \d B_v^H,
\label{fbm.mean}
\end{equation}
\begin{equation}
\Var[B^H_t|\mathcal{F}_s] = \Vert \textbf{1}_{[s,t]}(r) \Vert _{\kappa,T}^2 - \Vert \Psi(s,t,v) \textbf{1}_{[0,s]}(v) \Vert _{\kappa,T}^2, 
\label{fbm.var}
\end{equation}
where, for $v \in (0,s)$,
\begin{equation}
\Psi(s,t,v)=\frac{\sin(\pi \kappa)}{\pi}v^{-\kappa}(s-v)^{-\kappa}\int_s^t \frac{r^\kappa(r-s)^\kappa}{r-v}\, \d r,
\end{equation}	
for $v \in \{0,s\}$, $\Psi(s,t,v)=0$. See \cite{fink2013conditional}.

\end{remark}

\subsection{Stochastic processes related to the fOU process}
\label{sec.related process}
There are some important processes $Z_t$ that are related to the fOU process, i.e., they may be defined based on the fOU process, as follows,
\begin{equation}
Z_t=g(X_t)
\label{g}
\end{equation}
where $X_t$ is the fOU process and the function $g(\cdot)$ is invertible. 
With $z=g(x)$, the cumulative distribution functions (CDFs) of $Z$ and $X$ are related, as follows,
\begin{equation}
F_Z(z) \overset{def}{=}\mathbb{P}[Z\leq z]=\mathbb{P}[X\leq x]\overset{def}{=} F_X(x).
\end{equation}
By differentiation, we have the relation of their probability density functions (PDFs) as
\begin{equation}
f_Z(z) \overset{def}{=} \frac{\d F_Z(z)}{\d z}=\frac{\d F_X(x)}{\d x} \frac{\d x}{\d z} \overset{def}{=} \left[g^{\prime}(x)\right]^{-1}f_X(x).
\label{eq.pdf}
\end{equation}

Since the PDF of the fOU process is available, the PDF of a stochastic process related to the fOU process is also available, by means of \Cref{eq.pdf}. We are interested in the following three processes.
\begin{enumerate}[label=\arabic*),leftmargin=3em]
\item The geometric fractional Ornstein-Uhlenbeck (GfOU) process is defined as the process whose logarithm is the fOU process,
\begin{equation}
\left\{
\begin{array}{cll}
Z_t & = & e^{X_t} \\
\d X_t & = & \lambda(\mu-X_t)\,\d t + \sigma \, \d B^H_t \\
\end{array}
\right.
\label{GfOU}
\end{equation}
This process is found in finance, where we tend to model the logarithm of asset prices.

\item The fractional Cox-Ingersoll-Ross (fCIR) process is an extension of the classical Cox-Ingersoll-Ross (CIR) process, which is used for short-term interest rate and stochastic volatility modeling \cite{cox2005theory, heston1993closed}. Compared to the CIR process, the fCIR process has the advantage to model the ``memory phenomenon" in financial data \cite {mishura2018fractional}, like in rough volatility. It is shown in \cite{fink2011fractional} that, for the case $\kappa>0$, an fCIR process with zero mean,
\[
\d Z_t = -\lambda Z_t \, \d t + \sigma \sqrt{|Z_t|} \, \d B^H_t,
\]
is the square of the fOU process before hitting zero for the first time, i.e.,
\begin{equation}
\left\{
\begin{array}{cll}
Z_t & = & \frac{1}{4} \sigma ^2 X_t^2 \\
\d X_t & = & - \frac{\lambda}{2}X_t\,\d t + \d B^H_t\\
\end{array}
\right.
\label{fCIR}
\end{equation}

\item \Cref{fCIR} is an example of a polynomial process. Polynomial processes form a class of processes that are obtained by a polynomial map of the underlying process \cite{filipovic2016polynomial}. This class plays an increasingly important role in finance as it provides a tractable relationship between underlying factors and resulting prices. For example, a second-order polynomial process is employed for the valuation of electricity storage contracts in \cite{Ware2020, boonstra2021valuation}. Using the same increasing polynomial map construction method as in \cite{Ware2020}, we here consider a polynomial process with a third-order polynomial map, i.e.,
\begin{equation}
\dsp{
\left\{
\begin{array}{cll}
Z_t & = & \frac{1}{6} \delta X_t^3+\frac{1}{2} (1-\delta) X_t^2 \\
\d X_t & = & \lambda(\mu-X_t)\,\d t + \sigma \, \d B^H_t \\
\end{array}
\right.
}
\label{poly}
\end{equation}
where $\delta \in [0,1]$.

\end{enumerate}

\section{Conditional distribution of the fOU process}
\label{sec.3}
In the previous section, we reviewed the conditional distribution of the fOU process, which is a conditional normal distribution with expectation as in \Cref{fou.mean} and variance in \Cref{fou.var}. In this section, we calculate the relevant integrals by which the conditional distributions of the fOU process and several related processes can be derived for all relevant Hurst indices. 

With a discretization in the time interval $[0,s]$, i.e., $0=s_0\leq s_1 \leq \dots \leq s_n=s$, and the corresponding discrete realizations of $\B ^H$, according to the introduction of integrals with respect to fBm in \cite{pipiras2001classes}, the conditional expectation can be calculated by
\begin{equation}
\mathbb{E}[X_t|\mathcal{F}_s] \approx X_s \, e^{-\lambda(t-s)}+\mu(1-e^{-\lambda(t-s)})+ \sum_{i=0}^{n-1}\Psi_c(s,t,t_{i})(B^H_{t_{i+1}}- B^H_{t_{i}}).  
\label{fou.mean.appro}
\end{equation}

The variance, in \Cref{fou.var}, is expressed as the difference of two terms, each being the square of a norm that is not easily calculated, since the expression  involves a nontrivial integration with multiple  variables and singularities in its integral function. We first simplify the expression of the conditional variance in \Cref{sec.3.1} and obtain a solution for all $H\in(0,1)$ in \Cref{sec.3.2}, which is summarized in \Cref{algorithm_1}.

\subsection{Simplification of the conditional variance}
\label{sec.3.1}
Noting that both terms in \Cref{fou.var} contain indicator functions, we rewrite them as
\begin{equation}
\begin{aligned}
\Vert c(r) \textbf{1}_{[s,t]}(r) \Vert _{\kappa,T}^2 
&= A_{\kappa} \int_0^T z^{-2\kappa}\left(\int_z^T \frac{r^\kappa c(r) \textbf{1}_{[s,t]}(r)}{(r-z)^{1-\kappa}}\,\d r \right)^2\,\d z 
\\&=A_{\kappa} \left[\int_0^s z^{-2\kappa}\left(\int_s^t \frac{r^\kappa c(r)}{(r-z)^{1-\kappa}}\,\d r \right)^2\,\d z \right.
\\& \hspace{1cm} \left. +\int_s^t z^{-2\kappa}\left(\int_z^t \frac{r^\kappa c(r)}{(r-z)^{1-\kappa}}\,\d r \right)^2\,\d z\right]
\end{aligned}
\label{term1}
\end{equation}
and 
\begin{equation}
\begin{aligned}
\Vert \Psi_c(s,t,v) \textbf{1}_{[0,s]}(v) \Vert _{\kappa,T}^2 
&= A_{\kappa} \int_0^T z^{-2\kappa}\left(\int_z^T \frac{v^\kappa \Psi_c(s,t,v) \textbf{1}_{[0,s]}(v)}{(v-z)^{1-\kappa}}\,\d v \right)^2\,\d z 
\\ &=A_{\kappa} \int_0^s z^{-2\kappa} G^2(z)\,\d z,
\end{aligned}
\label{term2}
\end{equation}
respectively, where
\begin{equation}
\begin{aligned}
G(z)&=\int_z^s \frac{v^\kappa \Psi_c(s,t,v)}{(v-z)^{1-\kappa}}\,\d v \\
&=\frac{\sin(\pi \kappa)}{\pi} \int_z^s (v-z)^{\kappa-1} (s-v)^{-\kappa} \left(\int_s^t \frac{r^\kappa (r-s)^{\kappa}}{r-v}c(r)\,\d r \right)\,\d v.
\end{aligned}
\end{equation}

Using a new  integration variable, $u\in(0,\infty)$, with 
\begin{equation}
u=\frac{v-z}{s-v}\quad \Longrightarrow\quad v=\frac{us+z}{u+1},\quad \frac{\d v}{\d u}=\frac{s-z}{(1+u)^2}, \quad \frac{1}{v-z}=\frac{1+u}{u(s-z)},
\end{equation}
and interchanging the order of integration, yield
\begin{equation}
\begin{aligned}
G(z)&=\frac{\sin(\pi \kappa)}{\pi} \int_0^\infty \frac{u^{\kappa-1}}{1+u} \left(\int_s^t \frac{r^\kappa (r-s)^{\kappa}}{r-v}c(r)\,\d r \right) \,\d u
\\&=\frac{\sin(\pi \kappa)}{\pi}\int_s^t r^\kappa(r-s)^{\kappa-1} c(r) \int_0^\infty \frac{u^{\kappa-1}}{u+\frac{r-z}{r-s}}\,\d u \,\d r.
\end{aligned}
\end{equation}

Here, we recognise the Mellin transform (see \cite[Table~1.14.4] {olver2010nist})\footnote{http://dlmf.nist.gov/1.14.T4}
\begin{equation}
\int_0^\infty \frac{u^{\kappa-1}}{u+a}\,\d u=\frac{\pi}{\sin(\pi\kappa)}a^{\kappa-1},\quad 0<\kappa<1.
\label{Melli transform}
\end{equation}
This gives us,
\begin{equation}
G(z)=\int_s^t r^\kappa(r-z)^{\kappa-1} c(r)\,\d r.
\label{Gz}
\end{equation}
We have derived this relation for $0<\kappa<1$, as assumed in \Cref{Melli transform}, however, in later formulas we need this expression, or a similar result, also for $\kappa \le0$. For this, we use the classical technique of {\em analytic continuation}. For details on this topic, we refer to \ref{appendix.analytic continuation}.

Substituting \Cref{Gz} into \Cref{term2} and comparing with \Cref{term1}, we find that $\Vert \Psi_c(s,t,v) \textbf{1}_{[0,s]}(v) \Vert _{\kappa,T}^2$ is  part of $\Vert c(r) \textbf{1}_{[s,t]}(r) \Vert _{\kappa,T}^2$ and the expression of the conditional variance in~\Cref{fou.var} can be simplified, from two terms to one term only:
\begin{equation}
\Var[X_t|\mathcal{F}_s] = A_{\kappa} \int_s^t z^{-2\kappa}\left(\int_z^t \frac{r^\kappa c(r)}{(r-z)^{1-\kappa}}\,\d r\right)^2\,\d z. 
\label{var}
\end{equation}

\subsection{Calculation of the conditional variance}
\label{sec.3.2}
Knowing that $$\Gamma(\kappa+1)=\kappa\Gamma(\kappa) \;\;\mbox{ and } \;\; \frac{\pi \kappa}{\sin(\pi\kappa)}=\Gamma(1-\kappa)\Gamma(1+\kappa),$$ the front factor $A_{\kappa}$ of \Cref{var} can also be written as 
\begin{equation}
A_{\kappa} =\frac{\kappa(1-4\kappa^2)\Gamma(1-\kappa)}{\Gamma(2-2\kappa)\Gamma(\kappa)}.
\end{equation}

To further illustrate how to calculate the variance, we write \Cref{var} in the form, 
\begin{equation}
\Var[X_t|\mathcal{F}_s] = \frac{ \Gamma(1-\kappa)}{\Gamma(2-2\kappa)\Gamma(\kappa+1)}H(s,t,\kappa),
\end{equation}
with
\begin{equation} 
H(s,t,\kappa) = (1-4\kappa ^2)\int_s^t z^{-2\kappa} h^2(\kappa,z)\,\d z, 
\label{Hz}
\end{equation}

\begin{equation}
h(\kappa,z) = \kappa \int_z^t \frac{r^\kappa c(r)}{(r-z)^{1-\kappa}}\,\d r,
\label{hz}
\end{equation}
and
\begin{equation}
c(r)=\sigma e^{-\lambda(t-r)}.
\label{cr}
\end{equation}

\subsubsection{Calculation of \texorpdfstring{$H(s,t,\kappa)$}{H(s,t,k}}
A set of transformations is now proposed for the calculation of $H(s,t,\kappa)$ in \Cref{Hz}:
\begin{enumerate}[label=\arabic*),leftmargin=3em]
	\item  First, we use the transformation $x=z^{1-2\kappa}$, and obtain
	\begin{equation}
	H(s,t,\kappa)=(1+2\kappa)\int_{s^{1-2 \kappa}}^{t^{1-2 \kappa}} h^2(\kappa,z)\,\d x;
	\end{equation}
	
	\item Next, using $\displaystyle y = \frac{x-s^{1-2 \kappa}}{t^{1-2 \kappa}-s^{1-2 \kappa}}$ yields
	\begin{equation}
	H(s,t,\kappa)=(1+2\kappa)(t^{1-2 \kappa}-s^{1-2 \kappa})\int_{0}^{1} h^2(\kappa,z)\,\d y;
	\end{equation}
	
	\item The final transformation is given by $y=\frac{1}{2} \textrm{erfc}(-w)$, and we get
	\begin{equation}
	H(s,t,\kappa)=\frac{(1+2\kappa)(t^{1-2 \kappa}-s^{1-2 \kappa})}{\sqrt{\pi}}\int_{-\infty}^\infty e^{-w^2} h^2(\kappa,z)\,\d w,
	\label{final_inte}
	\end{equation}
	where $\textrm{erfc}(\cdot)$ is the complementary error function 
	\begin{equation}
	\textrm{erfc}(w)=\frac{2}{\sqrt{\pi}}\int_w^\infty e^{-t^2}\,\d t,
	\end{equation}
	with the following properties
	\begin{equation}
	\textrm{erfc}(-\infty)=2,\quad \textrm{erfc}(\infty)=0,\quad \frac{\d y}{\d w}=\frac{1}{\sqrt{\pi}} e^{-w^2}.
	\end{equation}	
\end{enumerate}

The final integral in \Cref{final_inte} can be approximated efficiently by the Trapezoidal Rule, 
\begin{equation}
\int_{-\infty}^\infty e^{-w^2} f(w)\,\d w\approx m\sum_{n=-\infty}^\infty e^{-(mn)^2}f(mn),
\end{equation}
where $m>0$ is the step size. Because of the fast convergence of the series, the range of summation can be cut off to a smaller interval $t\in[-a,a]$, $a>0$:
\begin{equation}
\int_{-\infty}^\infty e^{-w^2} f(w)\,\d w\approx m\sum_{|mn|\leq a} e^{-(mn)^2}f(mn).
\label{trapezoidal rule}
\end{equation}
For details about the Trapezoidal Rule, see \cite[\S5.4]{gil2007numerical}. In our case, $f(w)=h^2(\kappa,z)$, with $w$ a function of $z$.

\subsubsection{Calculation of \texorpdfstring{ $h(\kappa,z)$}{h(k,z}}
Here we will derive convenient computable expressions for the $h(\kappa,z)$ term from \Cref{hz}, which is the only involved part remaining.

We expand $c(r)$ from \Cref{cr} in powers of $r$ and get an expansion with fast convergence, i.e.,
\begin{equation}\label{eq:G02}
c(r)=\sum_{n=0}^\infty c_n r^n, \quad c_n=\sigma e^{-\lambda t}\, \frac{\lambda^n}{n!}.
\end{equation}
Then, we obtain
\begin{equation}\label{eq:G03}
h(\kappa,z)=\sum_{n=0}^\infty c_nR_n(\kappa,z), \quad R_n(\kappa,z)=\kappa\int_z^t r^{\kappa+n}(r-z)^{\kappa-1}\,\d r.
\end{equation}
Initially, we consider $\kappa > 0$, because for $r=z$ we need convergence of the integral. Integrating by parts gives us,
\begin{equation}\label{eq:G04}
 R_n(\kappa,z)=\int_z^t r^{\kappa+n}\,\d (r-z)^{\kappa}=
t^{\kappa+n}(t-z)^\kappa-
 (\kappa+n)\int_z^t r^{\kappa+n-1}(r-z)^\kappa\,\d r.
\end{equation}
In this representation, we can accept $\kappa >-1$, so the integration by parts gives us again an analytic continuation\footnote{For details on analytic continuation, we refer again to \ref{appendix.analytic continuation}.} of $R_n(\kappa,z)$  with respect to $\kappa$.

By using the fact that $(r-z)^\kappa=(r-z)^{\kappa-1}(r-z)$, a simple recursion formula follows
\begin{equation}\label{eq:G06}
R_{n}=\frac{\kappa t^{\kappa+n}(t-z)^\kappa+z(\kappa+n)R_{n-1}}{2\kappa+n},\quad n\ge1.
\end{equation}

The  functions $R_{n}$ can be expressed in terms of Gauss hypergeometric functions. For details about these functions, see \cite{Olde:2010:GHF}. In particular, we have the integral representation and power series,
\begin{equation}\label{eq:G07}
\begin{array}{r@{\,}c@{\,}l}
\FG{\alpha}{\beta}{\gamma}{\zeta}&=&\dsp{\frac{\Gamma(\gamma)}{\Gamma(\beta)\Gamma(\gamma-\beta)}\int_0^1 t^{\beta-1}(1-t)^{\gamma-\beta-1}(1-\zeta t)^{-\alpha}\,\d t,}\\[1.5ex]
&=&\dsp{\sum_{n=0}^\infty \frac{(\alpha)_n(\beta)_n}{n!\,(\gamma)_n} \zeta ^n,\quad (\alpha)_n=\frac{\Gamma(\alpha+n)}{\Gamma(\alpha)}.}
\end{array}
\end{equation}
For the integral we assume that  $\Re \{\gamma\}>\Re \{\beta\}>0$, where $\Re \{\cdot\} $ means the real part of the input argument, and that $\zeta$ is a complex number not  in  the interval $[1,\infty)$, and for the series we need the conditions $\vert \zeta \vert <1$ and $\gamma \ne0,-1,-2,\ldots$.

Again, initially we assume that $\kappa > 0$ and  use the transformation $r=z/(1-u)$, to obtain
\begin{equation}\label{eq:G08}
R_n(\kappa,z)=\kappa z^{2\kappa+n}\int_0^{1-z/t}u^{\kappa-1}(1-u)^{-n-2\kappa-1}\,\d u.
\end{equation}
In this integral, we take $u=(1-z/t)v$, which gives us,
by invoking the integral in \eqref{eq:G07},
\begin{equation}\label{eq:G09}
\begin{array}{r@{\,}c@{\,}l}
R_n(\kappa,z)&=&\dsp{\kappa z^{2\kappa+n}(1-\frac{z}{t})^\kappa\int_0^1v^{\kappa-1}\left[1-\left(1-\frac{z}{t}\right)v\right]^{-n-2\kappa-1}\d v}\\
&=&
\dsp{z^{2\kappa+n}(1-\frac{z}{t})^\kappa \FG{n+2\kappa+1}{\kappa}{\kappa+1}{1-\frac{z}{t}}}.
\end{array}
 \end{equation}
The representation in the second line of \Cref{eq:G09} is valid for $\kappa>-1$, and it gives us again the analytic continuation to the domain of $h(\kappa,z)$ initially defined for $\kappa>0$.

\begin{remark}From \Cref{eq:G03}, we have observed that we do not need to handle a singularity at the endpoint $z=s$ when $s>0$. However, this will be needed when $s=0$, in which case the variable of integration, $z$  in \Cref{Hz}, will reach the value $z=0$, but the ${}_2F_1$-function in \Cref{eq:G09} is not defined at $z=0$. For this case, a different form of $R_n(\kappa,z)$ will be required, which is given by
\begin{equation}
R_n(\kappa,z)=\frac{\Gamma(\kappa+1)\Gamma(n+1+\kappa)}{2\cos (\pi\kappa)\Gamma(n+1+2\kappa)}z^{2\kappa+n}
+\frac{\kappa (t-z)^{\kappa} t^{n+\kappa}}{n+2\kappa}\FG{-\kappa-n}{1}{1-n-2\kappa}{\frac{z}{t}}.
\end{equation}
The way this form is derived is presented in \ref{appendix.s0}. The ${}_2F_1$-function here is not defined for  $z\to t$ when $\kappa>0$. In that case, \Cref{eq:G09} should be used. 
\end{remark}

\begin{algorithm}
\caption{Calculation of the conditional variance of the fOU process}
\label{algorithm_1}
  \KwIn{Parameters of fOU process $\lambda$, $\mu$, $\sigma$, $\kappa$; time points $s$, $t$; step size $m$ and summation range $a$ for Trapezoidal Rule in \Cref{trapezoidal rule}; the number of expansion terms $N$ in \Cref{eq:G03}.}
  \KwOut{The conditional variance $\Var$.}
  \eIf{$\kappa=0$}{
  \lIf{$\lambda=0$}{$\Var=\sigma^2(t-s)$}
  \lElse{
  $\Var=\dsp{ \frac{\sigma^2 (1-e^{-2\lambda})}{2\lambda}}$
  }
  }{
  \lIf{$s=t$}{$\Var=0$}
  \uElse{
  $I=0$, $M=\lfloor a/m \rfloor$\;
  \For{$i=-M$ to $M$}{
  $w=m\cdot i$, $\dsp{y=\frac{1}{2}\textrm{erfc}(-w)}$, $\dsp{x=(t^{1-2\kappa}-s^{1-2\kappa})y+s^{1-2\kappa}}$, $\dsp{z=x^{\frac{1}{1-2\kappa}}}$\;
  $h=0$\;
  \For{$j=0$ to $N$}{
  $c_n=\dsp{\sigma e^{-\lambda t}} \frac{\lambda^n}{n!}$\;
  \uIf{$n=0$}{
  \uIf{$s=0$ {\bf and} $z/t<1/2$}{
  $R_0=\dsp{\frac{\Gamma(\kappa+1)\Gamma(1+\kappa)}{2\cos (\pi\kappa)\Gamma(1+2\kappa)}z^{2\kappa}
+\frac{\kappa (t-z)^{\kappa} t^{\kappa}}{2\kappa}\FG{-\kappa}{1}{1-2\kappa}{\frac{z}{t}}}$}
\uElse{
  $R_0=\dsp{z^{2\kappa}(1-\frac{z}{t})^\kappa \FG{2\kappa+1}{\kappa}{\kappa+1}{1-\frac{z}{t}}}$} 
  }
  \uElse{
  $R_{n}=\dsp{\frac{\kappa t^{\kappa+n}(t-z)^\kappa+z(\kappa+n)R_{n-1}}{2\kappa+n}}$}
  $h \leftarrow h+c_n R_{n}$\;
  }
  $\dsp{I\leftarrow I+me^{-(m\cdot i)^2}h^2}$}
  $\Var=\dsp{\frac{ \Gamma(1-\kappa)}{\Gamma(2-2\kappa)\Gamma(\kappa+1)}\frac{(1+2\kappa)(t^{1-2 \kappa}-s^{1-2 \kappa})}{\sqrt{\pi}}I}$}}
\end{algorithm}

\section{Option pricing methodology}
\label{sec.option pricing}
In this section, we will consider the computation of conditional expectations based on the calculation of the expectation and variance of the fOU process described in the previous sections. 

Usually, option valuation starts from the risk-neutral valuation formula, which is a discounted expectation under the risk-neutral asset price process.
Risk neutrality is related to complete markets, no arbitrage and the asset price process being a martingale, i.e., the expected fair value of the discounted price process at a future time point is the present asset price. Processes based on fractional Brownian motion are not \sm martingales, so the risk-neutral pricing formula does not apply in this case.
Financial option valuation related to weather~\cite{benth2003arbitrage, brody2002dynamical}, or the asset price's volatility, is based on a non-tradable asset for which the real world probability measure is often employed. Under the assumption that such a measure has been defined, option pricing can be cast in an integral form: 
\begin{equation}
    V(t,x)=e^{-r\Delta t} \, \mathbb{E}[V(T,y)|\mathcal{F}_{t}]=e^{-r\Delta t} \int_{-\infty}^{\infty} V(T,y) f(y|T,\mathcal{F}_t) \,\d y,
\label{valuation formula COS integral form}
\end{equation}
where $\Delta t:=T-t$, $V$ denotes the option value, $x$ the state variable at time $t$ and $y$ at time $T$, $f(y|T,\mathcal{F}_t)$ the transition probability density, and $r$ the interest rate.

Here, we focus on the European option pricing problem when the asset prices are modeled as the fOU related processes, as described in \Cref{sec.related process}. For these processes, the transition probability, $f(y|T,\mathcal{F}_t)$ in \Cref{valuation formula COS integral form}, is the PDF of a normal distribution with expectation $\mu_y = \mathbb{E}[X_T|\mathcal{F}_t] $ and variance $\sigma^2_y = \Var[X_T|\mathcal{F}_t]$, as  in \Cref{fou.mean} and \Cref{fou.var}, respectively. 
It is therefore possible to obtain a closed-form solution of \Cref{valuation formula COS integral form} for some specific processes. For example, if the asset price follows a GfOU process as \Cref{GfOU}, and the payoff reads
\begin{equation}
V(T,y) = \max \left[ \eta \left( e^y -K \right), 0 \right] \quad \mbox{with} \quad \eta =\left\{ 
\begin{array}{cl}
1 & \mbox{for a call},\\
-1 & \mbox{for a put},\\
\end{array}
\right.
\end{equation}
where $K$ is the strike price, the closed-form solution of \Cref{valuation formula COS integral form} for a call option can be derived as 
\begin{equation}
V(t,x)=e^{-r\Delta t+\frac{1}{2}\sigma^2_y+\mu_y}\left[1 \!-\! \Phi\left(\frac{\displaystyle \log K \!-\! \sigma^2_y \!-\! \mu_y}{\displaystyle\sigma_y}\right)\right] \!-\! K e^{-r\Delta t}\left[1 \!-\! \Phi\left(\frac{\displaystyle \log K \!-\! \mu_y}{\displaystyle \sigma_y}\right)\right],
\label{GfOU_pricing}
\end{equation}
and for a put option,
\begin{equation}
V(t,x)=K e^{-r\Delta t} \Phi\left(\frac{\displaystyle \log K-\mu_y}{\displaystyle \sigma_y}\right) - e^{-r\Delta t+\frac{1}{2}\sigma^2_y+\mu_y} \Phi\left(\frac{\displaystyle \log K-\sigma^2_y-\mu_y}{\displaystyle \sigma_y}\right).
\end{equation}
Here, $\Phi(\cdot)$ denotes the CDF of the standard normal distribution.

For more general processes, e.g., polynomial processes, a closed-form option value expression is difficult or even impossible to obtain. It is however possible to employ the COS method in such cases.
\subsection{The COS method}
The COS method, \cite{fang2009novel, fang2009pricing}, can be applied to compute expectations of (underlying) processes for which the characteristic function is available. For Markov processes with affine dynamics, the characteristic function can be easily derived by solving the Ricatti differential equations \cite{duffie2000transform}. Affinity is not invariant under polynomial transformations \cite{filipovic2020polynomial} and therefore the characteristic function of a stochastic model which is based on a polynomial process, $Z_t=g(X_t)$, generally does not exist. However, by using a transformation $z=g(x)$ in the definition of the COS method, the state variables can be conveniently chosen so that the characteristic function of the underlying process $X_t$, 
\begin{equation}
\dsp{\phi(u|T,\mathcal{F}_t) = e^{i \mu_y u - \frac{1}{2}\sigma^2_y u^2}},
\end{equation}
can be employed within the pricing formula, with, again, $\mu_y = \mathbb{E}[X_T|\mathcal{F}_t] $ and $\sigma^2_y = \Var[X_T|\mathcal{F}_t]$, as  in \Cref{fou.mean} and \Cref{fou.var}.

The conditional probability density function in \Cref{valuation formula COS integral form} is approximated by a truncated Fourier cosine expansion, which uses the characteristic function to recover the Fourier coefficients, as follows \cite{fang2009novel}:
\begin{equation}
    f(y|T,\mathcal{F}_t)\approx \frac{2}{d-b} \sum_{l=0}^{L-1} {}^{'}  \Re \left\{ {\phi}\left(\frac{l\pi}{d-b}\bigg|T,\mathcal{F}_t \right)  \cdot e^{-il\pi \frac{b}{d-b} } \right\}  \cos\left(l\pi \frac{y-b}{d-b}\right),
    \label{approximation density fourier cosine}
\end{equation}
where $[b,d]$ is the truncated integration interval, $L$ is the number of terms in the Fourier-cosine expansion and $\sum '$ implies that the first term of the summation is multiplied by $\frac{1}{2}$. 

By replacing the conditional density function $f(y|T,\mathcal{F}_t)$ in \Cref{valuation formula COS integral form} by its Fourier cosine expansion approximation \Cref{approximation density fourier cosine} and interchanging integration and summation, with the help of Fubini's theorem, the COS formula is obtained:
\begin{equation}
    V(t,x)\approx e^{-r\Delta t} \sum_{l=0}^{L-1} {}^{'} \Re\left\{ {\phi}\left( \frac{l\pi}{d-b}\bigg|T,\mathcal{F}_t \right) e^{il\pi \frac{-b}{d-b} } \right\} {\cal V}_l,
    \label{approximation valuation formula}
\end{equation}
where coefficients ${\cal V}_l$ are defined by:
\begin{equation}
    {\cal V}_l=\frac{2}{d-b}  \int_{b}^{d} V(T,y)  \cos\left(l\pi \frac{y-b}{d-b}\right) \d y,
    \label{coef V_k european COS}
\end{equation}
where, importantly, {\em the state variables $x$ and $y$ can be any function of respectively the asset prices $Z_{t}$ and $Z_T$}, e.g., in our case, $x=g^{-1}(Z_{t})=X_{t}$ and $y=g^{-1}(Z_{T})=X_{T}$, with $g(\cdot)$, invertible, as in \Cref{g}. This means that we can simply work with the characteristic function of the fOU process $X_t$, but price options under the more involved process $Z_t$. A closed-form solution of the coefficients ${\cal V}_l$ is available for various payoff functions and several choices of the state variables $x$ and $y$ \cite{fang2009novel}. 

\section{Numerical results}
\label{sec.5}
We present several numerical results for the methods presented, starting with the calculation of the conditional variances of the fBm and fOU processes, followed by the computation of probability density functions of the related processes, i.e., the GfOU, fCIR and polynomial processes, and some option pricing results using the COS method. These results show the accuracy and efficiency of the proposed computational techniques.

\subsection{Calculation of variance}
First of all, we focus on the accurate representation of the conditional variance by means of the integrals that were derived in Section~\ref{sec.3}.
\subsubsection{Choice of hyperparameters}
\begin{figure}[!htb]
	\centering
	\begin{subfigure}{0.54\textwidth}
	\includegraphics[width=\textwidth]{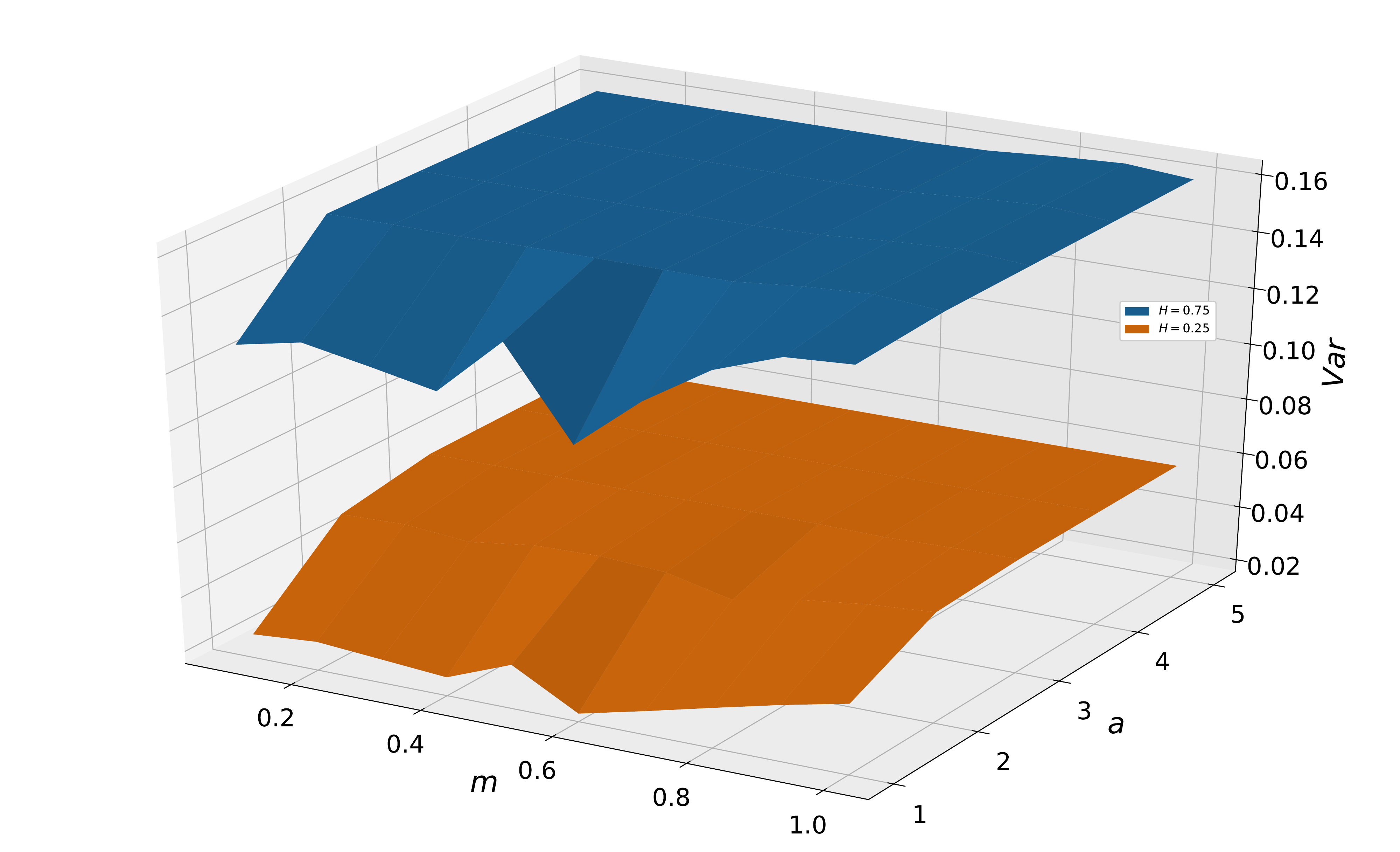}
	\caption{$m$ and $a$}
	\label{fig.convergence}
	\end{subfigure}
	\begin{subfigure}{0.42\textwidth}
	\includegraphics[width=\textwidth]{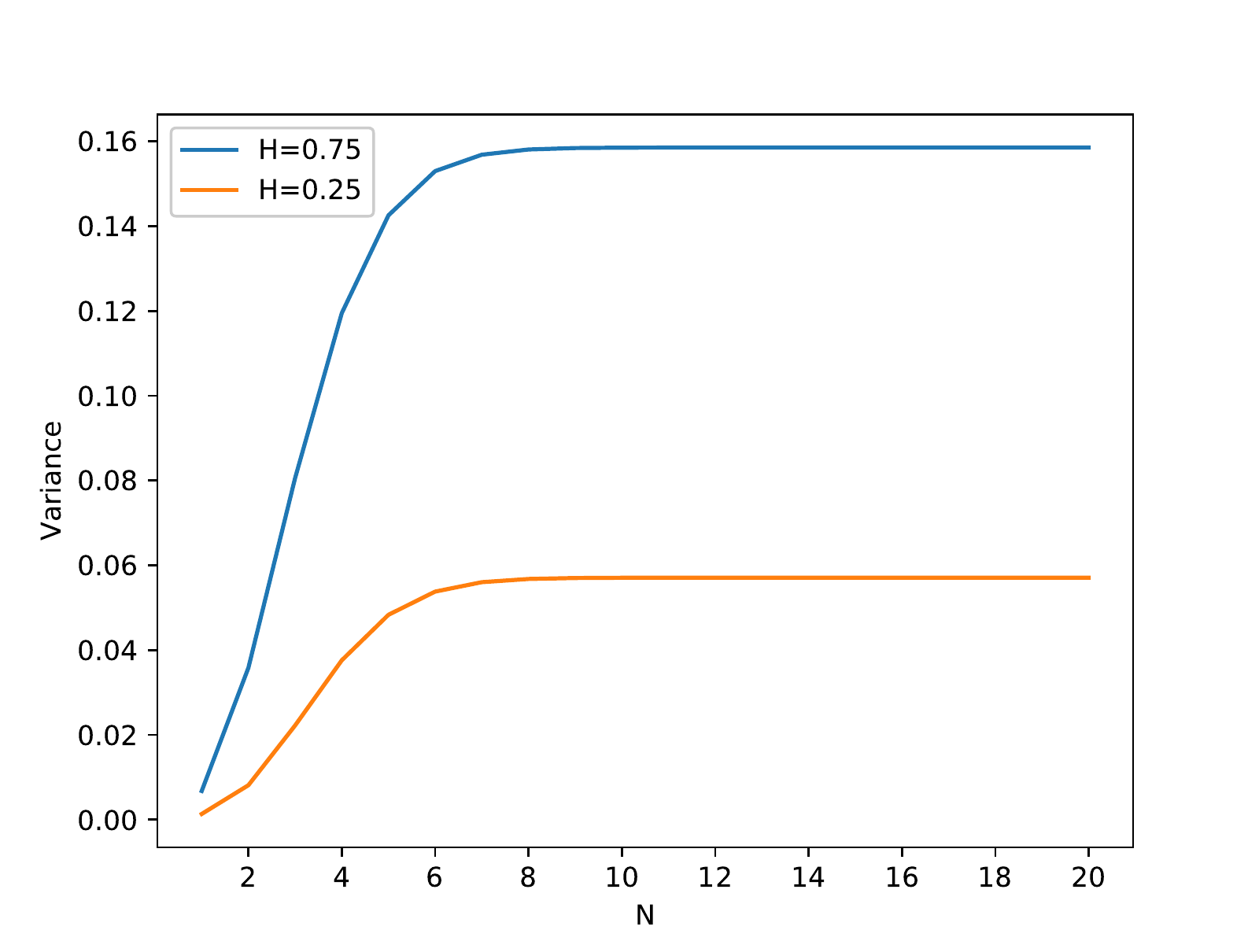}
	\caption{$N$}
	\label{fig.convergence_N}
	\end{subfigure}
	\caption{Calculated variances with $\sigma=0.3$, $\lambda=0.5$, $s=0$ and $t=5$ for two values of $H$. (a)  Step size $m$ varies from $0.1$ to $1.0$, and summation range $a$ from $1$ to $5$, with fixed $N=20$. (b) $N$ varies from $1$ to $20$, with $m=0.5$ and $a=5$ fixed. }
\end{figure}

As explained in the previous sections and also shown in \Cref{algorithm_1}, there are three hyperparameters in our numerical method to compute the integrals in the approximation of the conditional variance for the fOU process, i.e., the step size $m$, summation range $a$ for the Trapezoidal Rule in \Cref{trapezoidal rule}, and the number of expansion terms $N$ in \Cref{eq:G03}. To determine suitable values for these parameters, we perform the following experiment. For two values of $H$, and the value of $N$ sufficiently large (i.e. $N=20$), the computed results with varying $m$ and $a$ are presented in \Cref{fig.convergence}.  The figure shows that the summation range value $a$ has a bigger impact than the step size $m$. The numerical approximation has converged when $a$ is larger than $3$ and $m$ is smaller than $0.6$. We therefore set $a=5$ and $m=0.5$ in the subsequent numerical experiments.
Next, we consider the effect of the number of expansion terms $N$ on the computed results, also for the two values of $H$, and plot it in \Cref{fig.convergence_N}. It shows that our numerical scheme converges rapidly as $N$ increases, and setting N to be larger than $10$ already provides a reliable computational result. We may define a stopping criterion by checking when, e.g., $\vert c_n R_n / h \rvert $ is less than $10^{-8}$, then stop the summation. The results presented are representative for a large range of model parameters.

\subsubsection{Numerical performance}
To analyze the accuracy of our numerical technique, we compare the conditional ($s=3$) and unconditional ($s=0$) variances obtained by our method with Monte Carlo simulation. FBm and fOU, with $\sigma=0.3$, $\lambda=0.5$, are considered, with different Hurst indices $H$. For the simulation of fBm, we use a Python package named “fbm"\footnote{https://pypi.org/project/fbm/}. 

Two sets of parameters are considered, one with a step size of $0.01$ and the number of Monte Carlo paths of $10^4$, and another with the step size halved and the number of paths multiplied by 10. We report the relative errors in \Cref{table.var_error}, which are defined as
\begin{equation}
\epsilon = \frac{ \lvert \bar\sigma_1-\bar\sigma_2 \rvert}{\bar\sigma_1},
\end{equation}
where the standard deviation of the Monte Carlo results is $\bar\sigma_1$ and the computed standard deviation by numerical integration is $\bar\sigma_2$.

\begin{table}[!htb]
	\centering
	\caption{Conditional ($s=3$) and unconditional ($s=0$) standard deviations of fBm and the fOU process, with $\sigma=0.3$, $\lambda=0.5$, $t=s+5$ and Hurst index $H$ from $0.1$ to $0.9$. The numbers in brackets are relative errors (\%) compared with Monte Carlo simulations with two configurations, i.e, time step $0.01$, number of paths $10^4$, and time step $0.005$, number of paths $10^5$, respectively. Errors reported are the averages of $10$ experiments.}
		\begin{tabular}{ccccc}
			\toprule
			\multirow{2}{*}{$H$} & \multicolumn{2}{c}{fBm} & \multicolumn{2}{c}{fOU}\\ 
			\cmidrule(lr){2-3}
			\cmidrule(lr){4-5}
			& $s=0$ & $s=3$ & $s=0$ & $s=3$\\
			\midrule
			$0.1$& 1.1662 (0.92, 0.75) & 0.9774 (0.54, 0.32) & 0.2186 (0.80, 0.22) & 0.2175 (0.82, 0.25)\\ 
			$0.2$& 1.3796 (0.59, 0.19) & 1.2546 (0.41, 0.15) & 0.2310 (0.72, 0.16) & 0.2296 (0.42, 0.16)\\
			$0.3$& 1.6207 (0.44, 0.18) & 1.5544 (0.60, 0.15) & 0.2482 (0.46, 0.18) & 0.2470 (0.40, 0.19)\\
			$0.4$& 1.9036 (0.65, 0.22) & 1.8832 (0.43, 0.12) & 0.2708 (0.66, 0.17) & 0.2702 (0.47, 0.23)\\
			$0.5$& 2.2361 (0.52, 0.18) & 2.2361 (0.63, 0.23) & 0.2990 (0.61, 0.17) & 0.2990 (0.76, 0.17)\\
			$0.6$& 2.6265 (0.99, 0.21) & 2.5924 (0.68, 0.14) & 0.3334 (0.80, 0.21) & 0.3317 (0.62, 0.16)\\
			$0.7$& 3.0852 (0.55, 0.19) & 2.9025 (0.55, 0.14) & 0.3746 (0.53, 0.18) & 0.3633 (0.53, 0.25)\\
			$0.8$& 3.6239 (0.31, 0.17) & 3.0555 (0.52, 0.23) & 0.4238 (0.67, 0.21) & 0.3808 (0.54, 0.19)\\
			$0.9$& 4.2568 (0.41, 0.11) & 2.7760 (0.50, 0.22) & 0.4822 (0.57, 0.13) & 0.3491 (0.61, 0.21)\\
			\bottomrule[1pt]
		\end{tabular}
		\label{table.var_error}

\end{table}

\begin{table}[!htb]
	\centering
	\caption{CPU times (sec.) using our computational technique and Monte Carlo simulations for the variances of fBm and the fOU process with three representative $H$ indices (i.e., $H=0.3,\,0.5,\,0.7$). The triples are CPU times of our computational technique, Monte Carlo simulations with time step $0.01$, number of paths $10^4$ and time step $0.005$, number of paths $10^5$, respectively. Results reported are the averages of $10$ experiments.}
		\begin{tabular}{ccccc}
			\toprule
			\multirow{2}{*}{$H$} & \multicolumn{2}{c}{fBm} & \multicolumn{2}{c}{fOU}\\ 
			\cmidrule(lr){2-3}
			\cmidrule(lr){4-5}
			 & $s=0$ & $s=3$ & $s=0$ & $s=3$\\
			\midrule
			$0.3$& 1.1e-3, 1.30, 20.05 & 2.4e-3, 2.13, 65.74  & 1.1e-3, 1.47, 21.47 & 2.4e-3, 2.37, 63.12\\
			$0.5$& 1.2e-5, 0.44, 8.58 & 6.7e-6, 0.58, 11.68 & 1.1e-5, 0.44, 8.50  & 7.7e-6, 0.54, 11.70\\
			$0.7$& 1.1e-3, 1.30, 20.27 & 2.4e-3, 2.12, 65.67 & 1.1e-3, 1.54, 21.42 & 2.6e-4, 2.15, 62.91\\
			\bottomrule[1pt]
		\end{tabular}
		\label{table.time}

\end{table}
As shown in \Cref{table.var_error}, our numerical technique provides highly accurate results very close to those with the Monte Carlo simulations, with all average relative errors less than $1\%$. By using smaller time step and simulating more paths within the Monte Carlo simulation, the differences are even smaller. The CPU times\footnote{The computer used has an Intel Core i7-8700K CPU. The code is written in Python 3.} are also compared in \Cref{table.time}. Clearly, a numerical integration technique requires much less time than a Monte Carlo sampling method, because Monte Carlo simulation requires the generation of many paths with a small time step to obtain reliable results.

\begin{figure}[!htb]
	\centering
	\begin{subfigure}{0.44\textwidth}
	    \includegraphics[width=\textwidth]{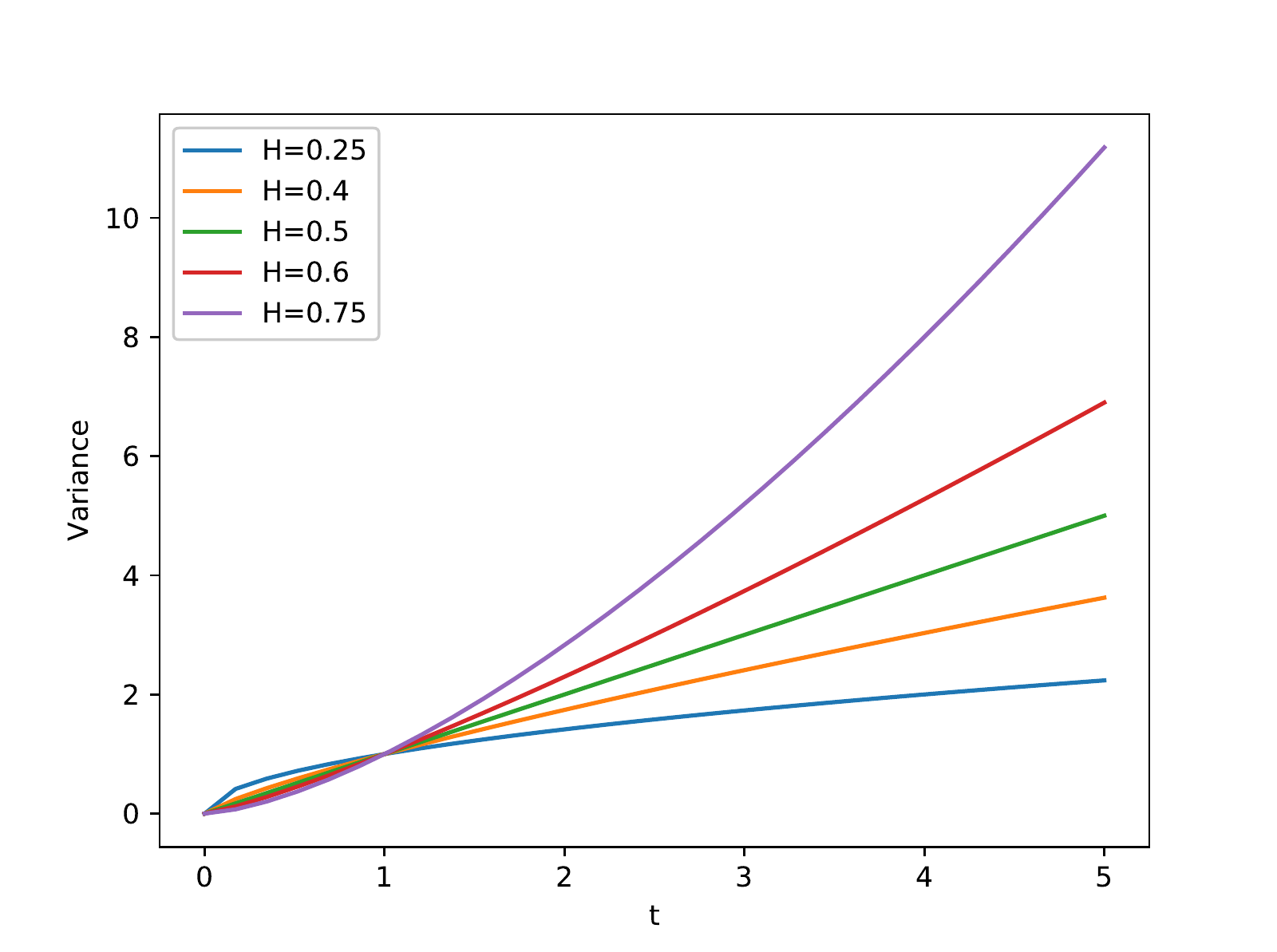}
	    \caption{$s=0$}
	\end{subfigure}
	\begin{subfigure}{0.44\textwidth}
	\includegraphics[width=\textwidth]{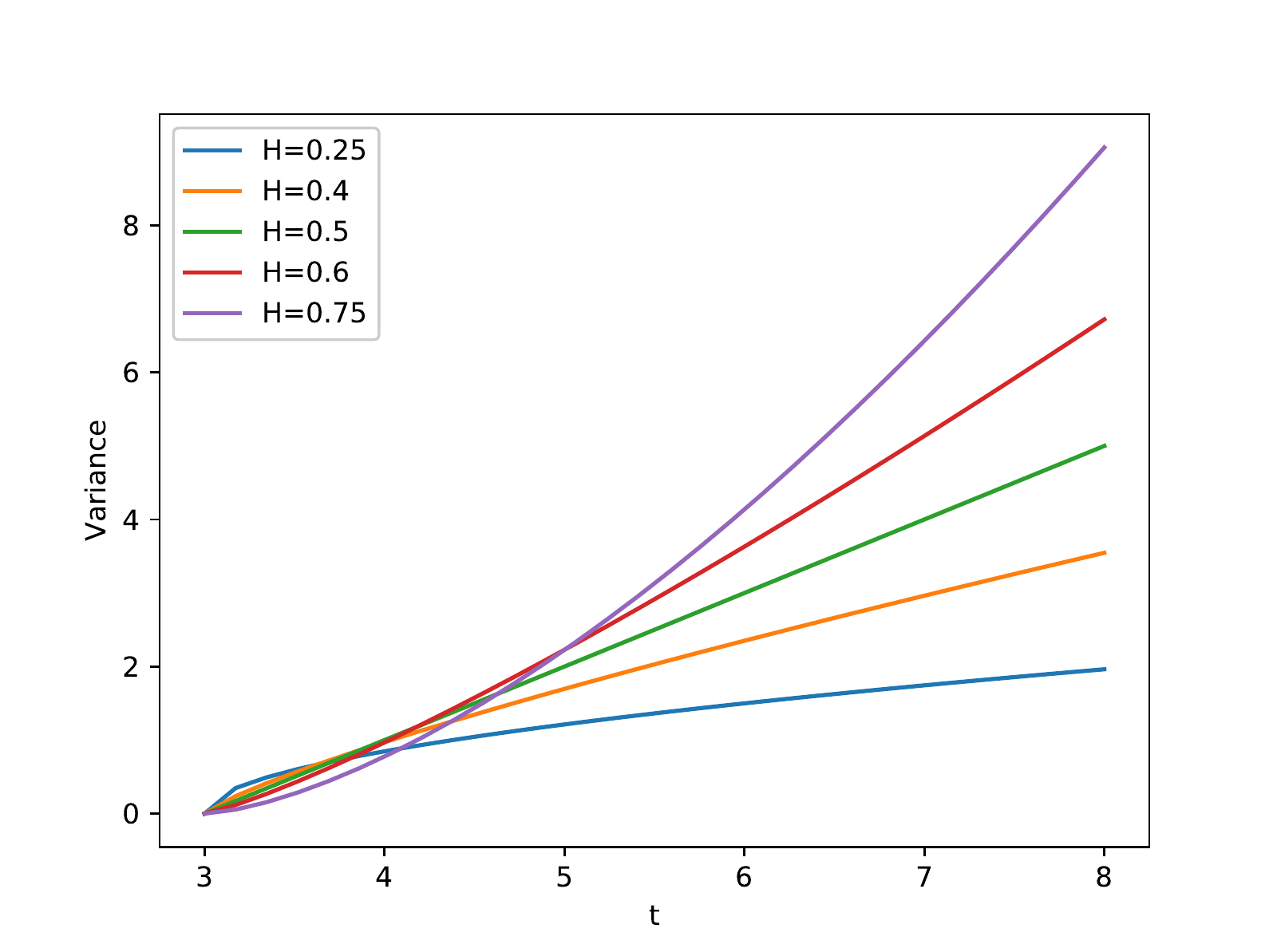}
	    \caption{$s=3$}
	\end{subfigure}	
	\caption{Variances of the fBm process for different $H$ and $t$ values.}
	\label{fig.var_fBm}
\end{figure}

\begin{figure}[htb]
	\centering
	\begin{subfigure}{0.44\linewidth}
	    \includegraphics[width=\textwidth]{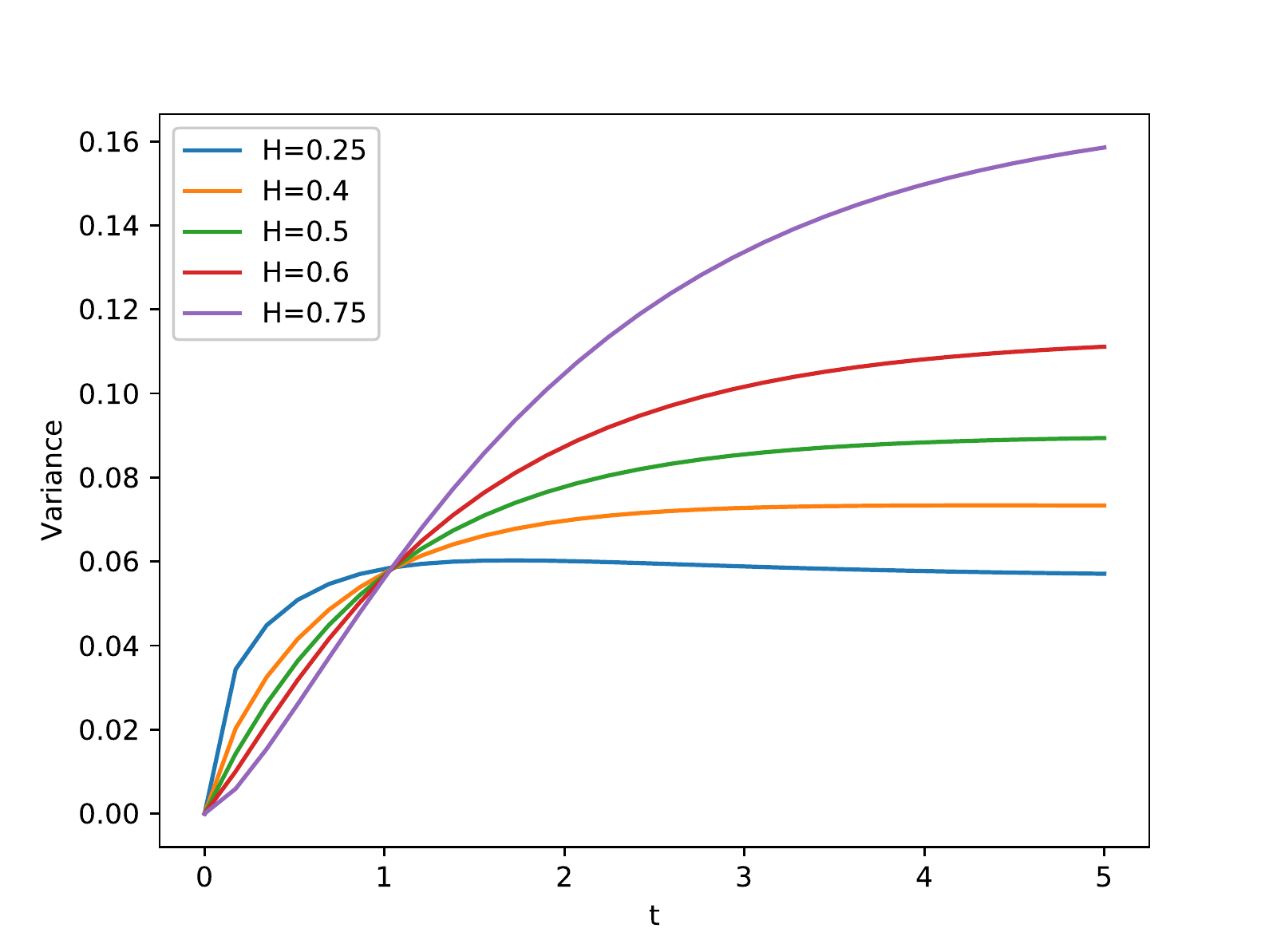}
		\caption{$s=0$}
	\end{subfigure}
	\begin{subfigure}{0.44\linewidth}
	\includegraphics[width=\textwidth]{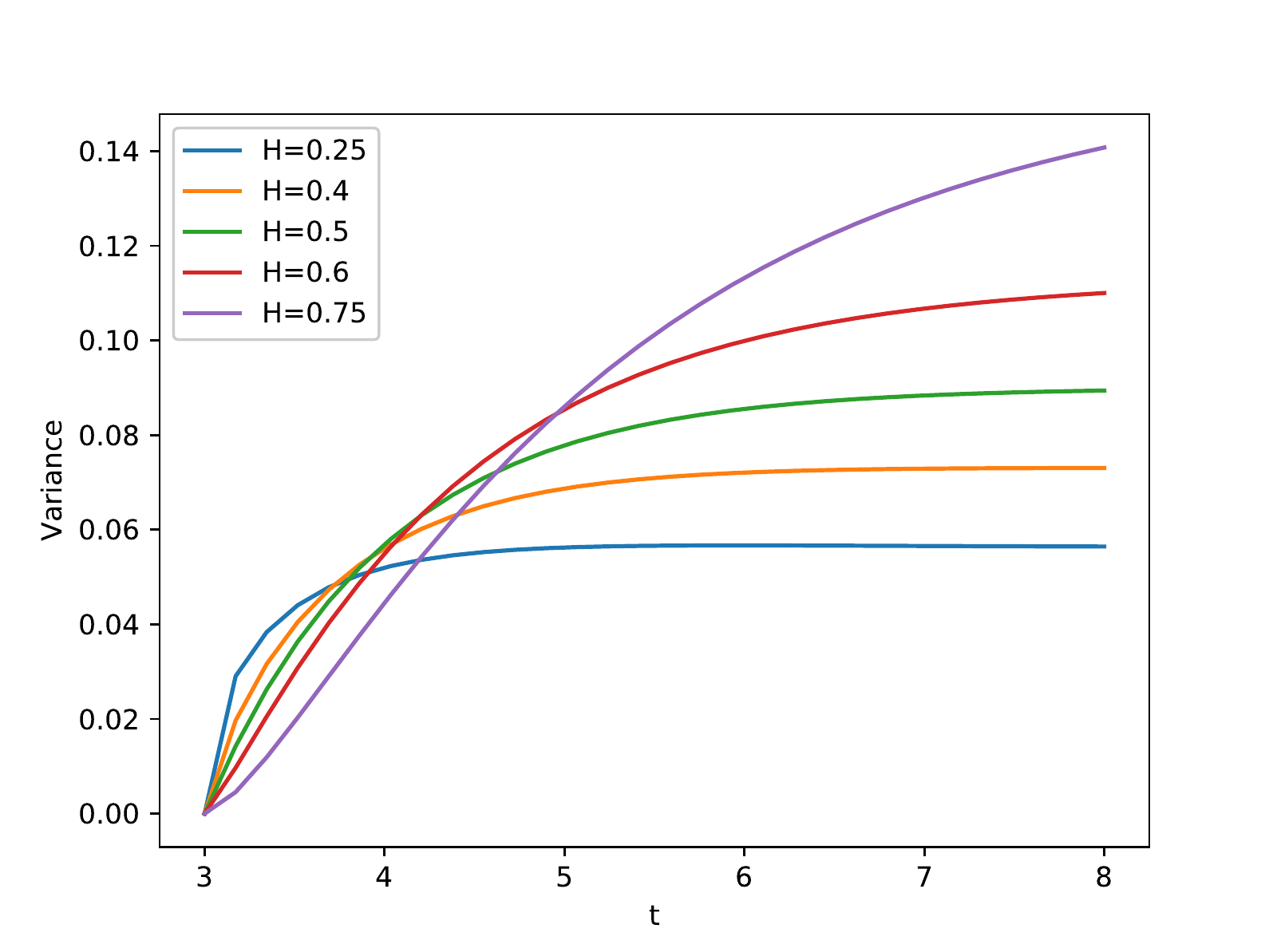}
	\caption{$s=3$}
	\end{subfigure}
	\caption{Variances of the fOU process for different $H$ and $t$, and $\sigma=0.3$, $\lambda=0.5$.}
	\label{fig.var_fOU}
\end{figure}

We further present the computed variances for different $H$, time point $s$ and with increasing time intervals (i.e., $t-s$) in \Cref{fig.var_fBm} (fBm) and \Cref{fig.var_fOU} (the fOU process with $\sigma=0.3$, $\lambda=0.5$). From these figures, we can conclude that our computational technique converges very well for all relevant $H$ indices, in a robust way. The curves for the different values of $H$ are very accurate, as compared to the literature~\cite{fink2013conditional}. As the time interval $t-s$ increases, the variance increases. With a sufficiently large time interval $t-s$, large values of the Hurst parameter $H$ results in higher variances. This makes sense as the Hurst index indicates the roughness of a process, with a higher value leading to smoother paths. Comparing the presented variances for $s=0$ and $s=3$,  different results are observed, for the same time interval $t-s$, which confirms that the variance is not a function $t-s$, but of $s$ and $t$ separately. 
	
\subsection{PDF recovery of fOU related processes}
As described in \Cref{sec.related process}, due to the relation between the PDFs of the fOU process and the related processes  in \Cref{eq.pdf}, we will compute the PDFs of the related processes by means of the PDF of the fOU process, which is normally distributed with the expectation in \Cref{fou.mean.appro} and the variance computed as in \Cref{algorithm_1}, and the transformation $g(\cdot)$ in~\Cref{g}.

\begin{figure}[!htb]
	\centering
	\begin{subfigure}[t]{0.30\textwidth}
    	\makebox[0pt]{\makebox[30pt][t]{\raisebox{40pt}{\rotatebox[origin=c]{90}{\scriptsize {GfOU}}}}}%
    	\includegraphics[width=\textwidth]{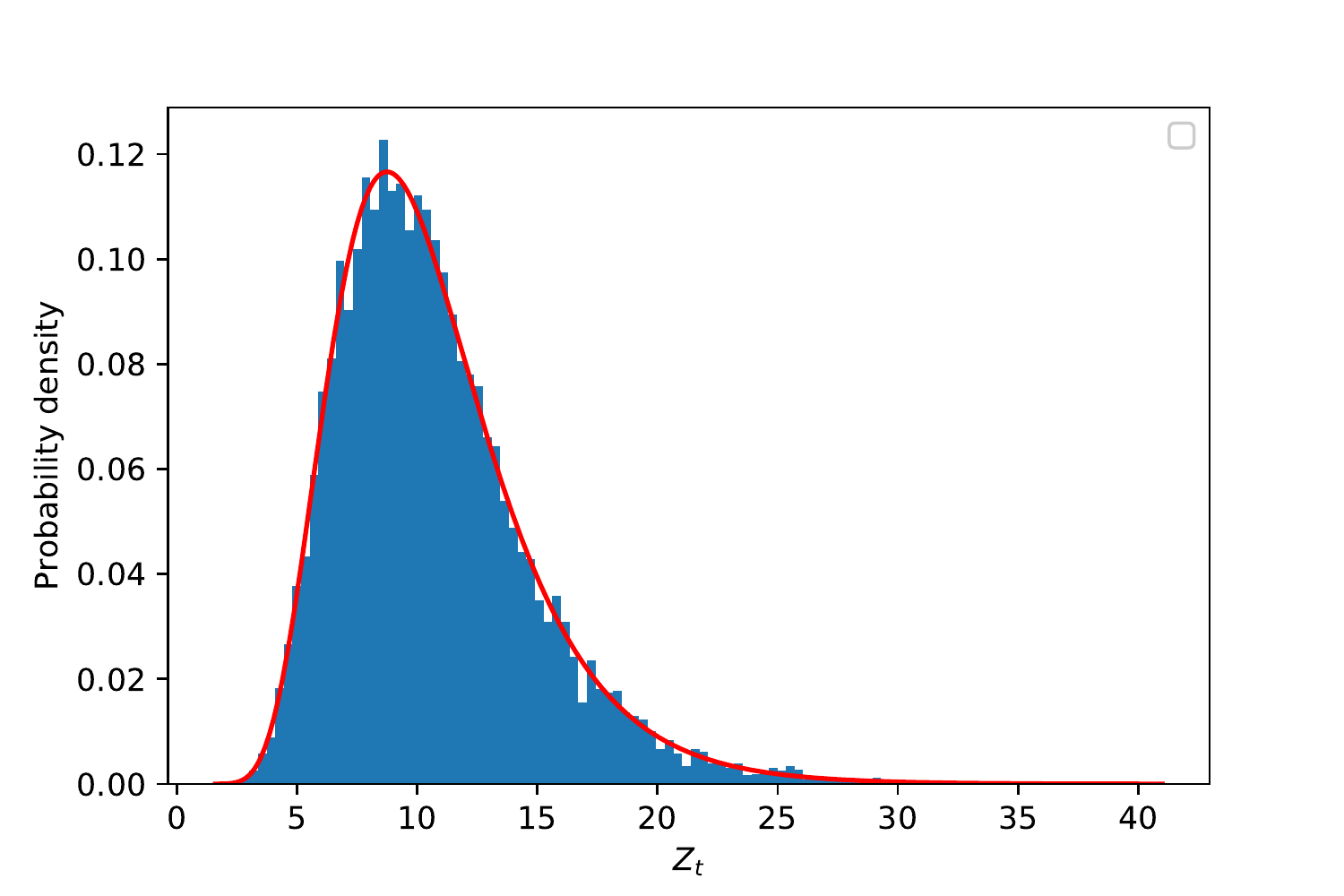}
    	\makebox[0pt]{\makebox[30pt][t]{\raisebox{40pt}{\rotatebox[origin=c]{90}{\scriptsize fCIR}}}}%
    	\includegraphics[width=\textwidth]{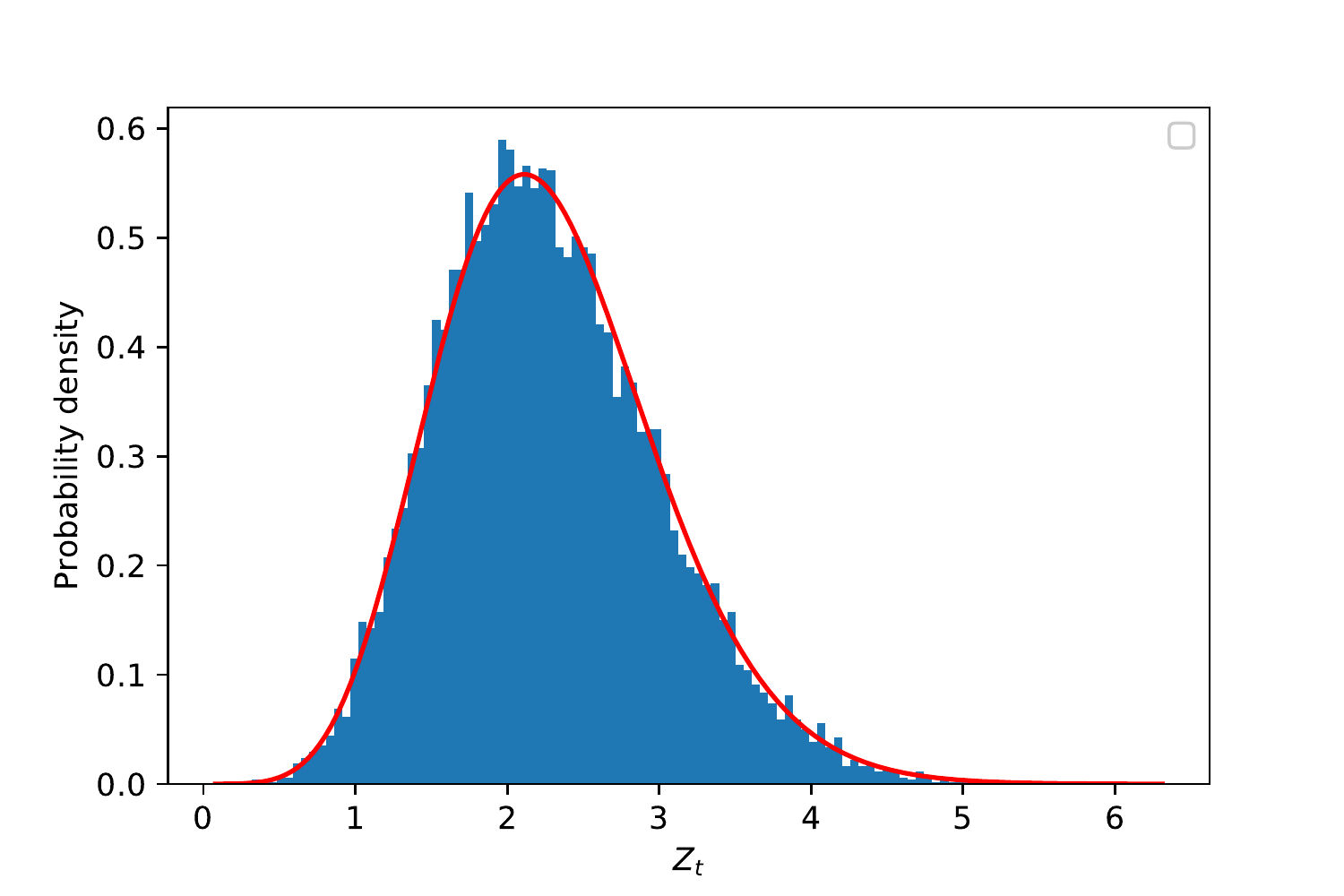}
    	\makebox[0pt]{\makebox[30pt][t]{\raisebox{40pt}{\rotatebox[origin=c]{90}{\scriptsize Polynomial}}}}%
    	\includegraphics[width=\textwidth]{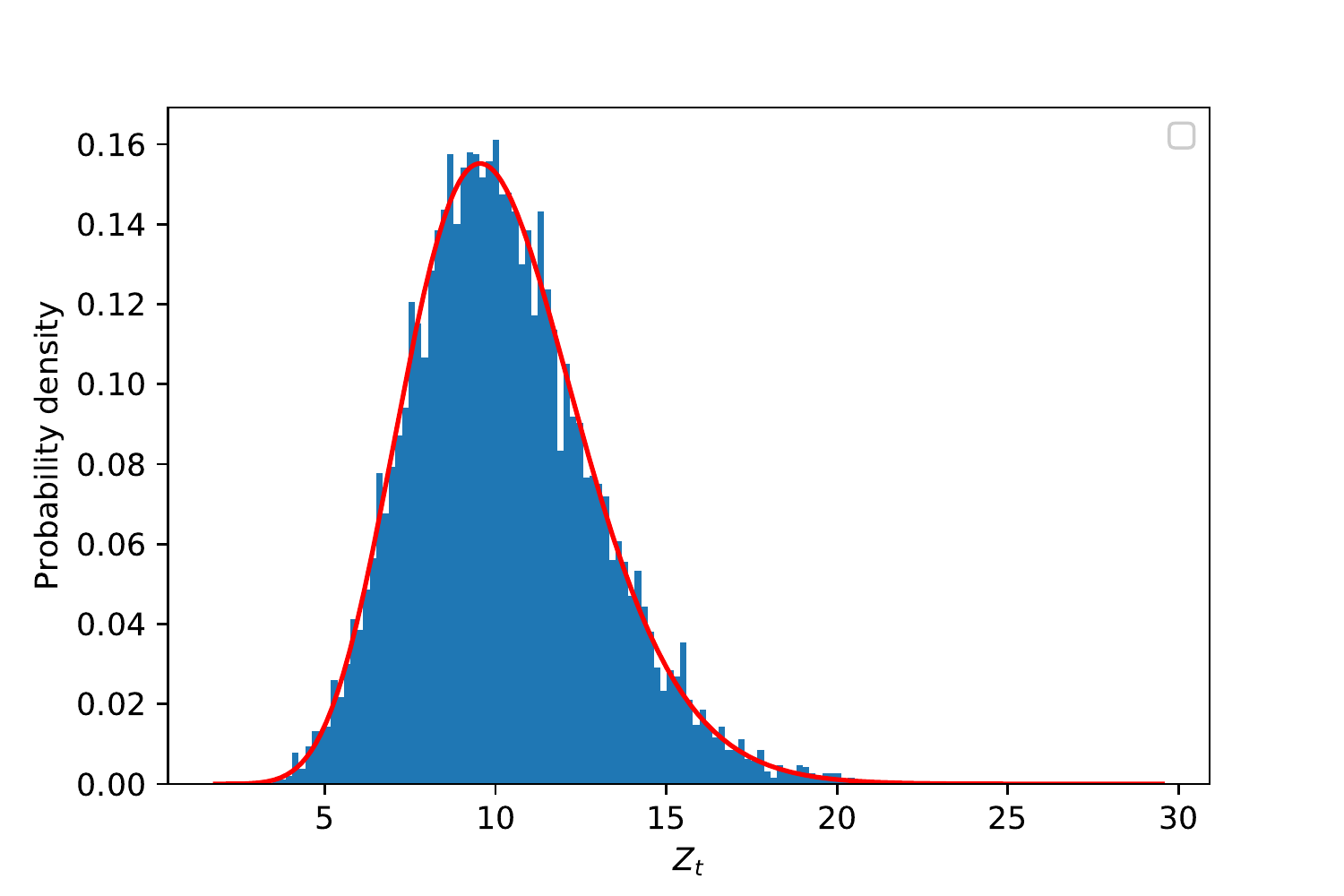}
    	\caption*{PDF with $s=0$, $t=3$}
    \end{subfigure}
	\begin{subfigure}[t]{0.30\textwidth}
	    \includegraphics[width=\textwidth]{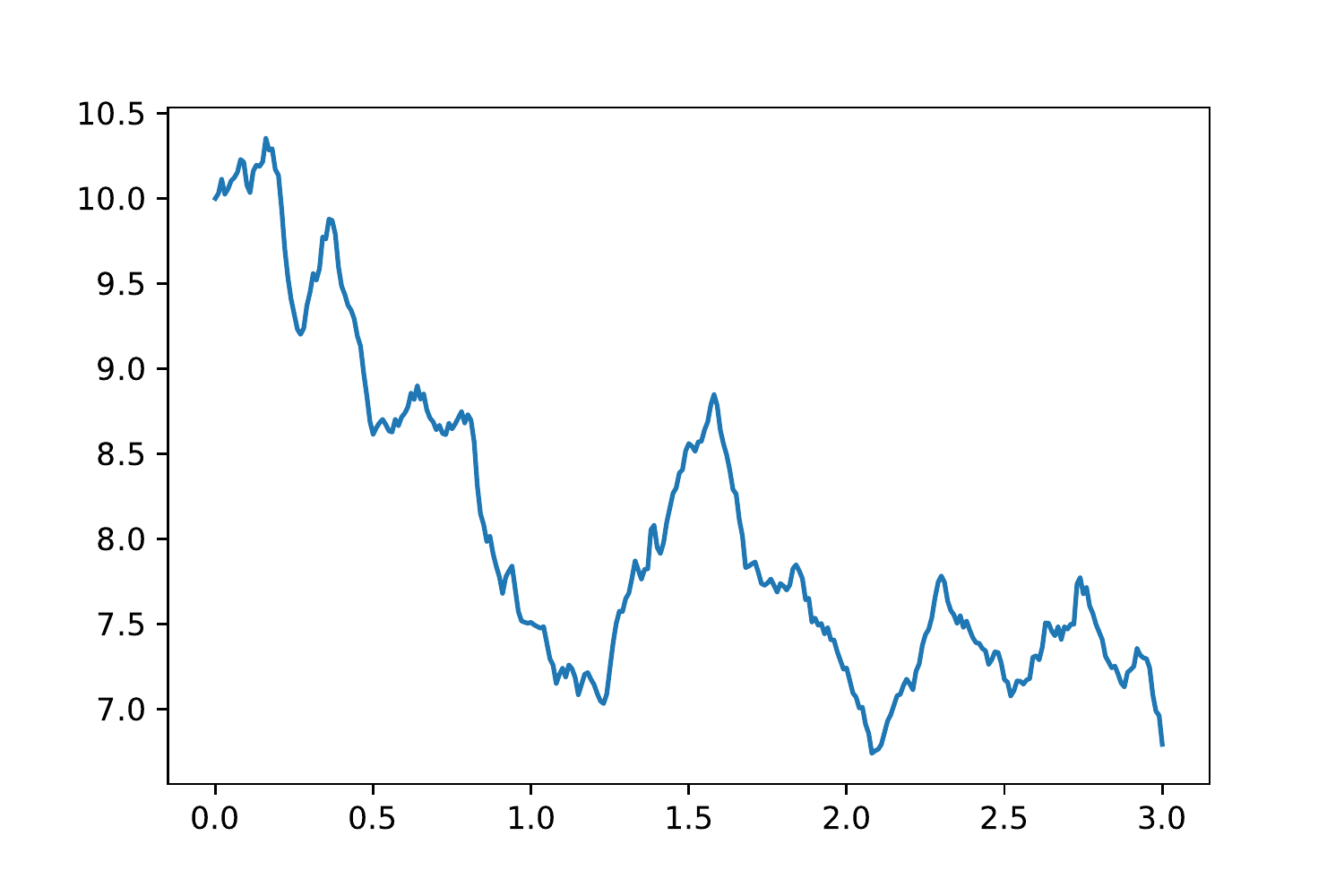}
	    \includegraphics[width=\textwidth]{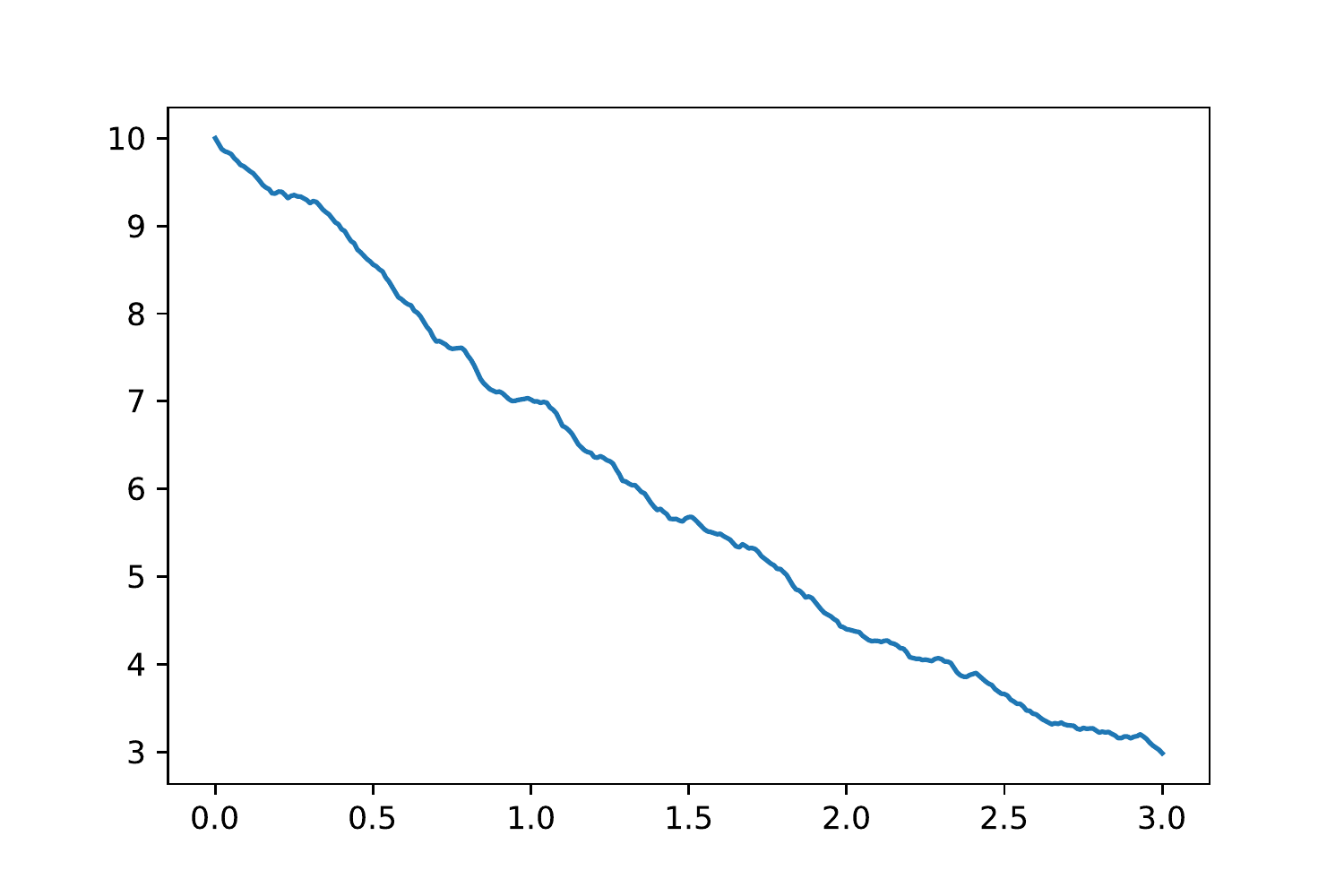}
	    \includegraphics[width=\textwidth]{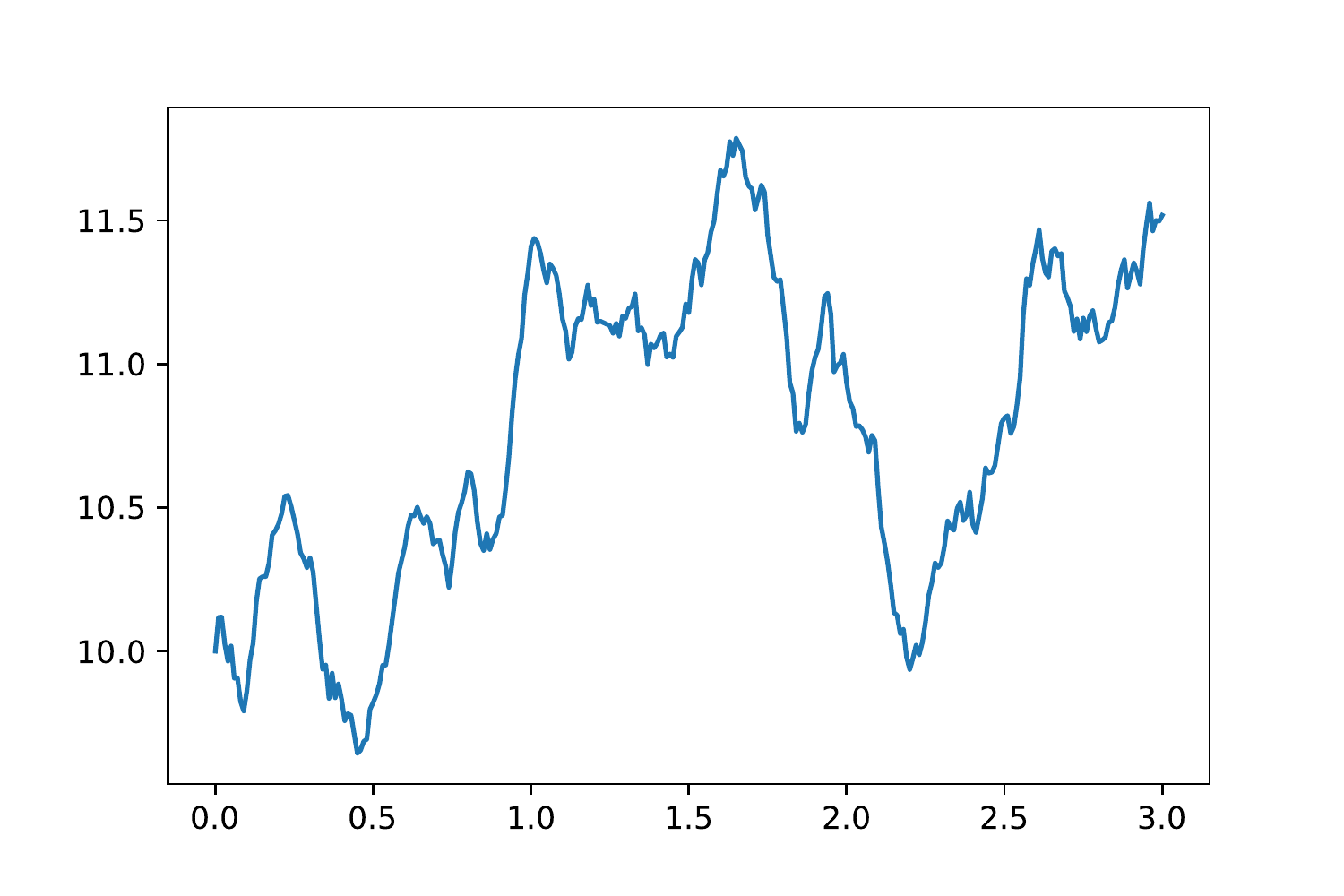}
	    \caption*{Historical path}
	\end{subfigure}
	\begin{subfigure}[t]{0.30\textwidth}
	    \includegraphics[width=\textwidth]{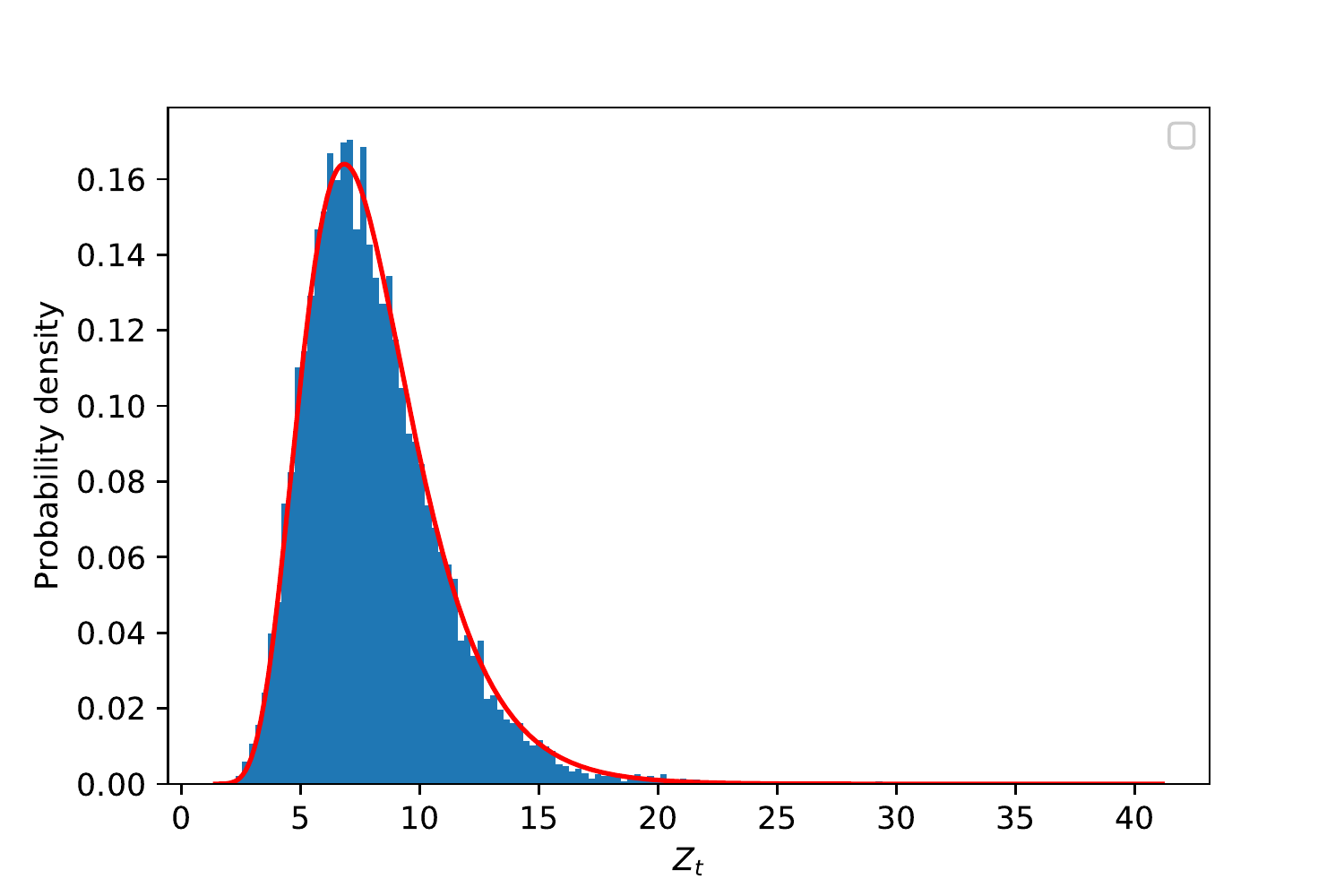}
	    \includegraphics[width=\textwidth]{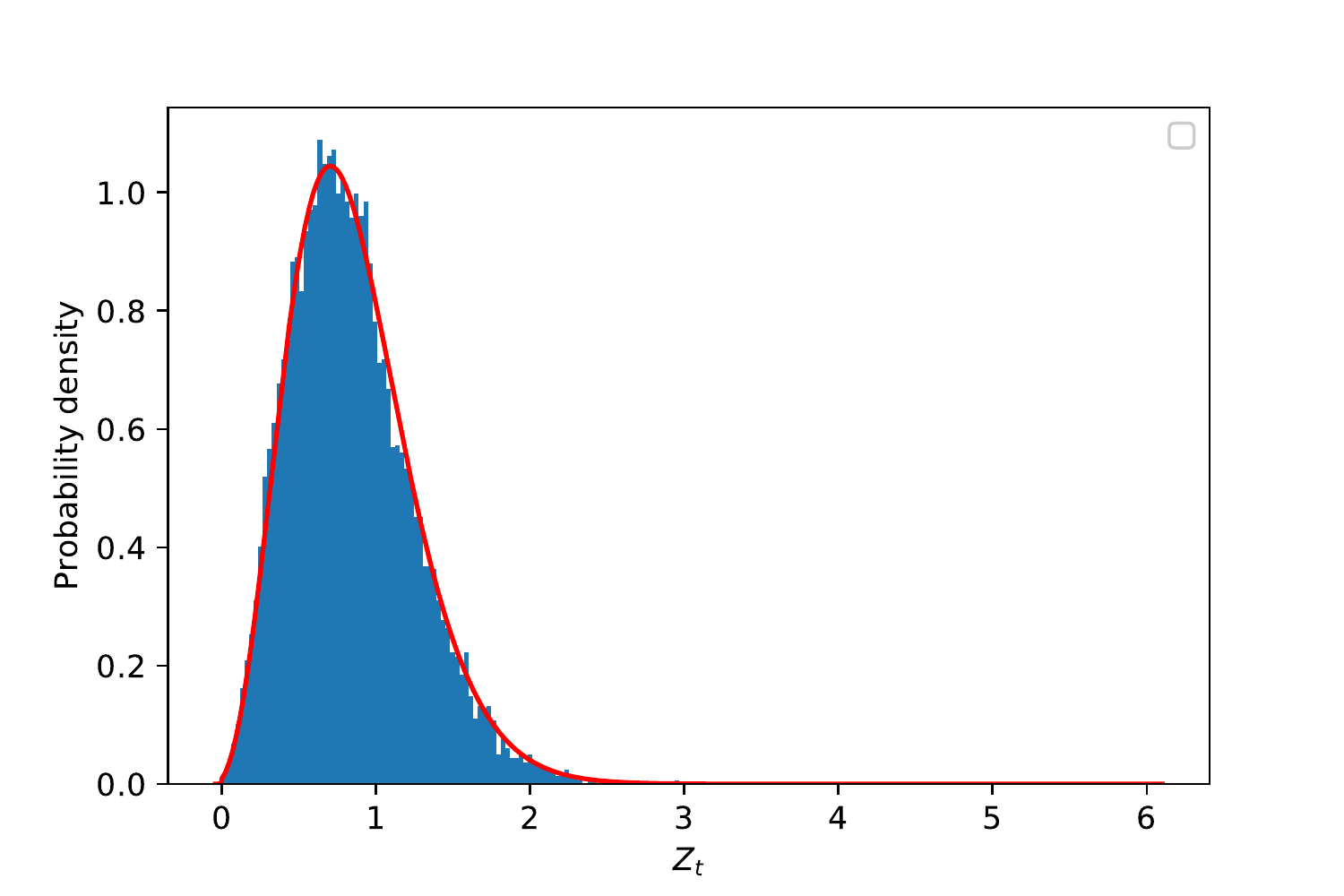}
	    \includegraphics[width=\textwidth]{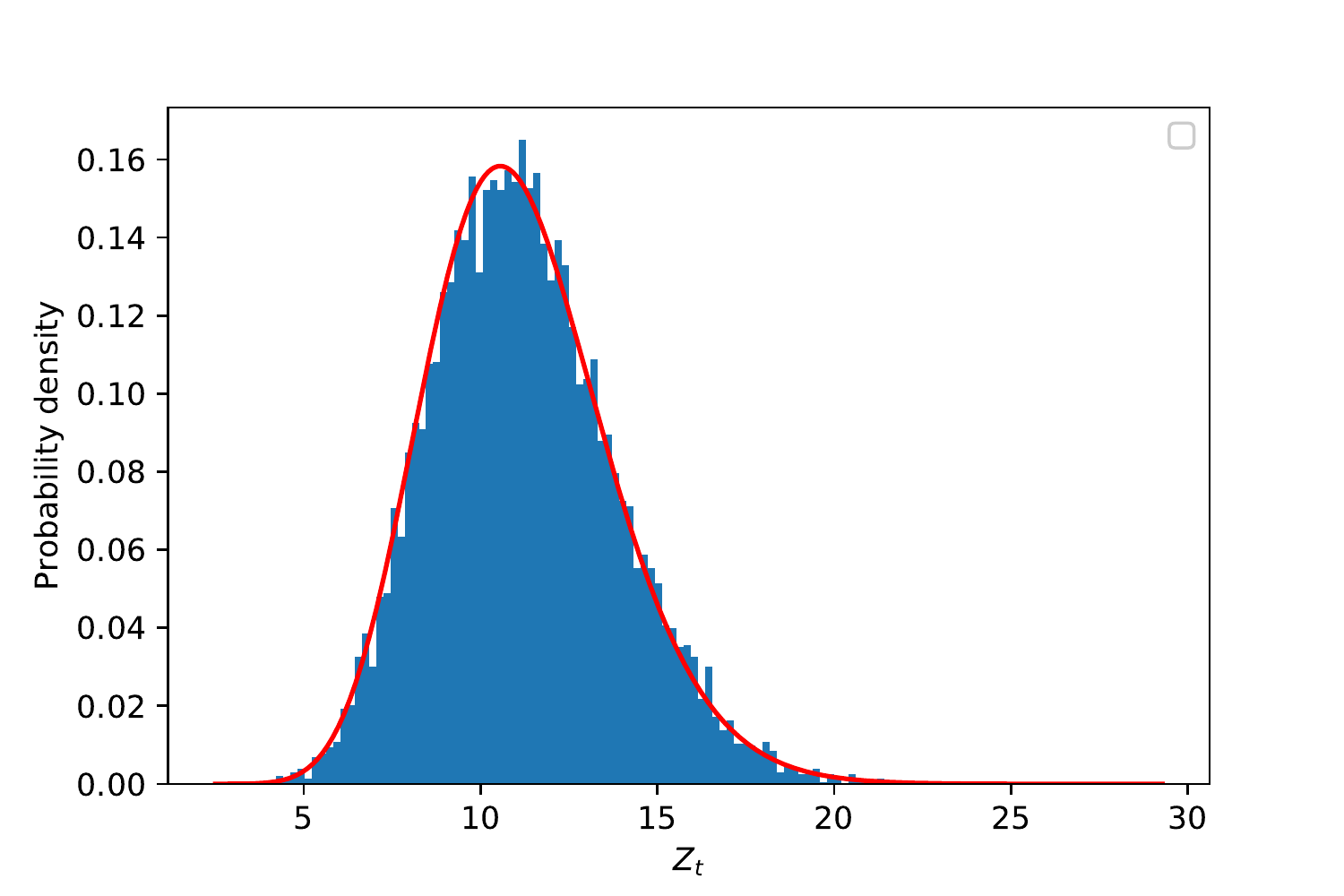}
	    \caption*{PDF with $s=3$, $t=6$}
	\end{subfigure}
	\caption{PDFs of the related processes with $H=0.75$, $\sigma=0.3$, $\lambda=0.5$, $Z_0=10$. The red curves are the PDFs obtained by our scheme, the blue histograms are Monte Carlo simulation results. From left to right, first column: PDFs at time $t=3$ with given initial value; second column: realized paths until time $s=3$; third column, PDFs at time $t=6$ on the condition of corresponding path in the second column. From top to bottom, first row: the GfOU process as in \Cref{GfOU} with $\mu = \textrm{log}(Z_0)$; second row: the fCIR process as in \Cref{fCIR}; third row: the polynomial process as in \Cref{poly} with $\delta = 0.8$ and $\mu=g^{-1}(Z_0)$, where $g(x)=\delta x^3/6+(1-\delta)x^2/2$.}
	\label{fig.pdf}
\end{figure}

\Cref{fig.pdf} plots the PDFs of the GfOU, fCIR and polynomial process, as in \Cref{GfOU}, \Cref{fCIR} and \Cref{poly}, respectively. We calculate the PDFs on the condition of either initial value ($s=0$, $t=3$) or realized historical paths ($s=3$, $t=6$), and compare them with the results by Monte Carlo simulation. It is shown that the calculated PDFs resemble the histograms of Monte Carlo simulation very well, which is an indication of the accuracy of our technique. Moreover, by comparing the first and third columns of \Cref{fig.pdf}, we can see the impact of the memory of the processes (the second column of \Cref{fig.pdf}) on the PDFs. 

\subsection{Option pricing}
The COS method enables us to price options if we have the characteristic function available for a function of the asset price process, like $X_t$ in $Z_t=g(X_t)$. When the asset prices are modeled as one of the fOU related processes, the COS method will be based on the characteristic function of the fOU process $X_t$. 

To test the performance of the COS method, we price a European call option under the GfOU process. This is just a reference example, since a closed-form solution is available in this case, as in \Cref{GfOU_pricing}. The errors of the COS method are presented in \Cref{table.cos_error} with the strike price $K=10$, the interval of integration $[b,d]=[\mu_y-10\sqrt{\sigma_y^2},\mu_y+10\sqrt{\sigma_y^2}]$ and a varying number of cosine terms $L$, see Equations~(\ref{approximation density fourier cosine}) and (\ref{approximation valuation formula}). It is shown that the COS method converges rapidly, as with only $L=16$ Fourier-cosine expansion terms, the errors are of the order $10^{-8}$.

\begin{figure}[!htb]
	\centering
	\begin{subfigure}{0.44\textwidth}
	    \includegraphics[width=\textwidth]{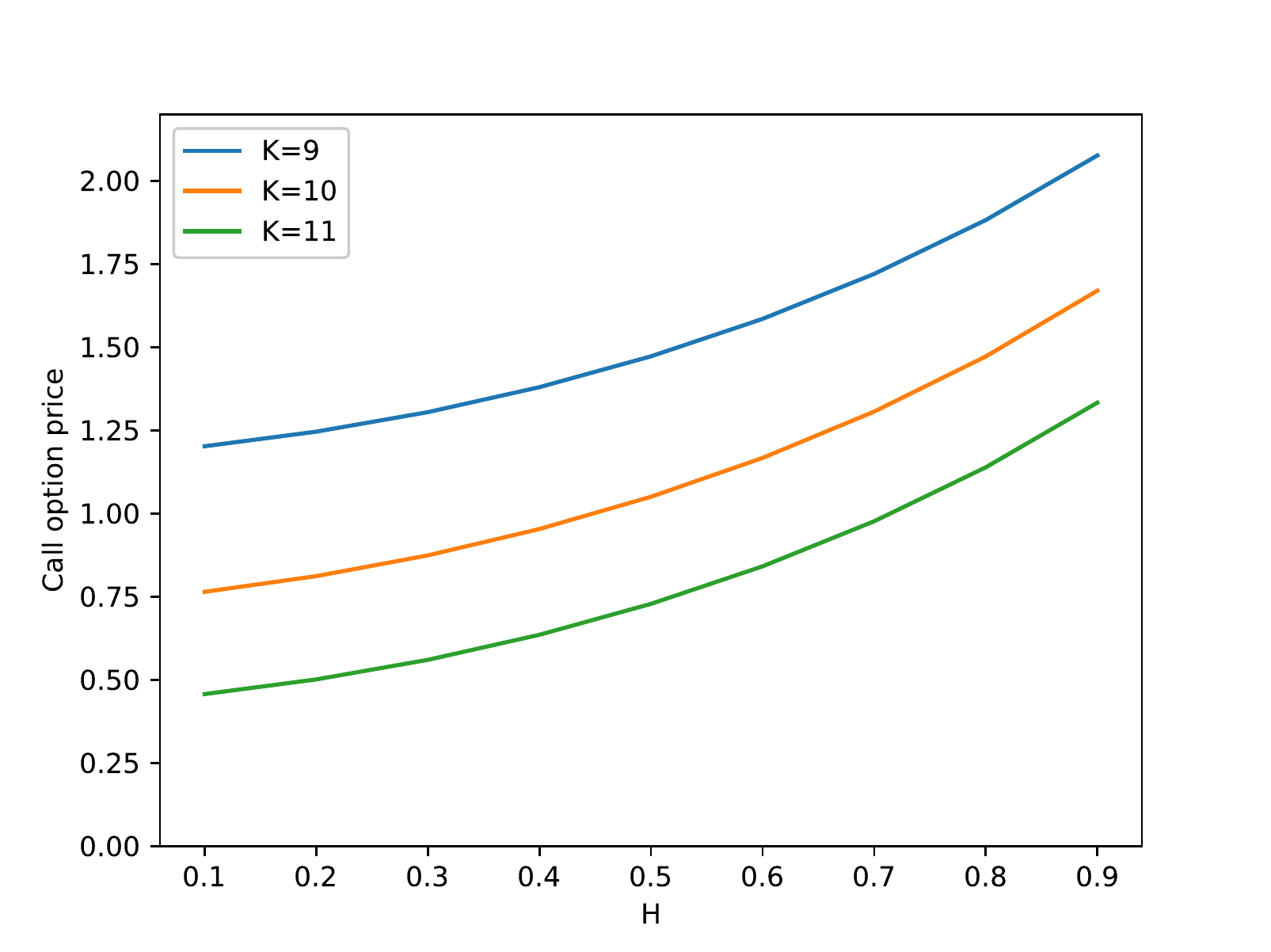}
	    \caption{Call option}
	\end{subfigure}
	\begin{subfigure}{0.44\textwidth}
	    \includegraphics[width=\textwidth]{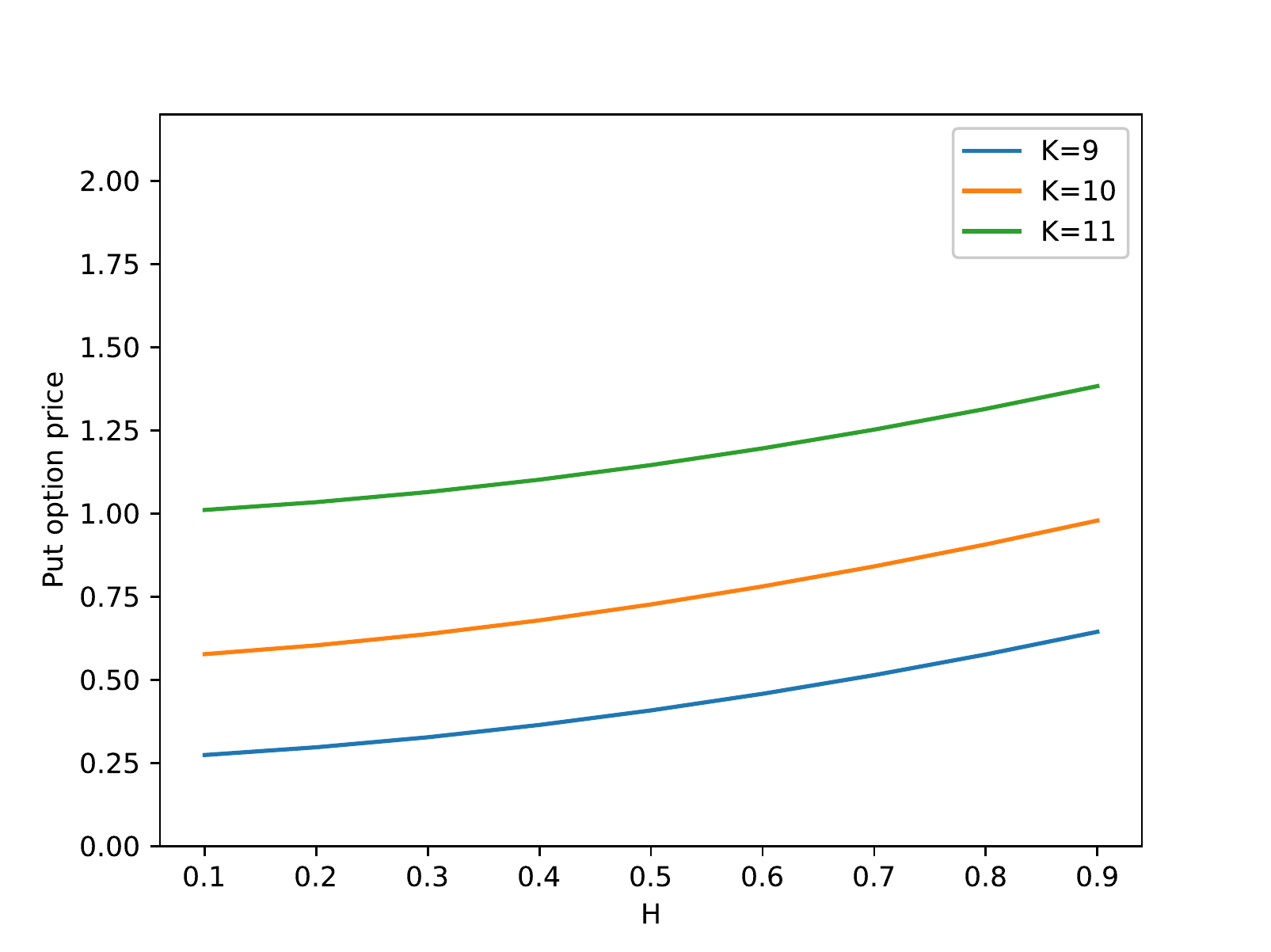}
	    \caption{Put option}
	\end{subfigure}
	\caption{European option values under the GfOU process with different $H$ and $K$, $t=0$ and $T=3$ fixed.}
	\label{fig.price_GfOU}
\end{figure}

\begin{figure}[htb]
	\centering
	\begin{subfigure}{0.44\textwidth}
	\includegraphics[width=\textwidth]{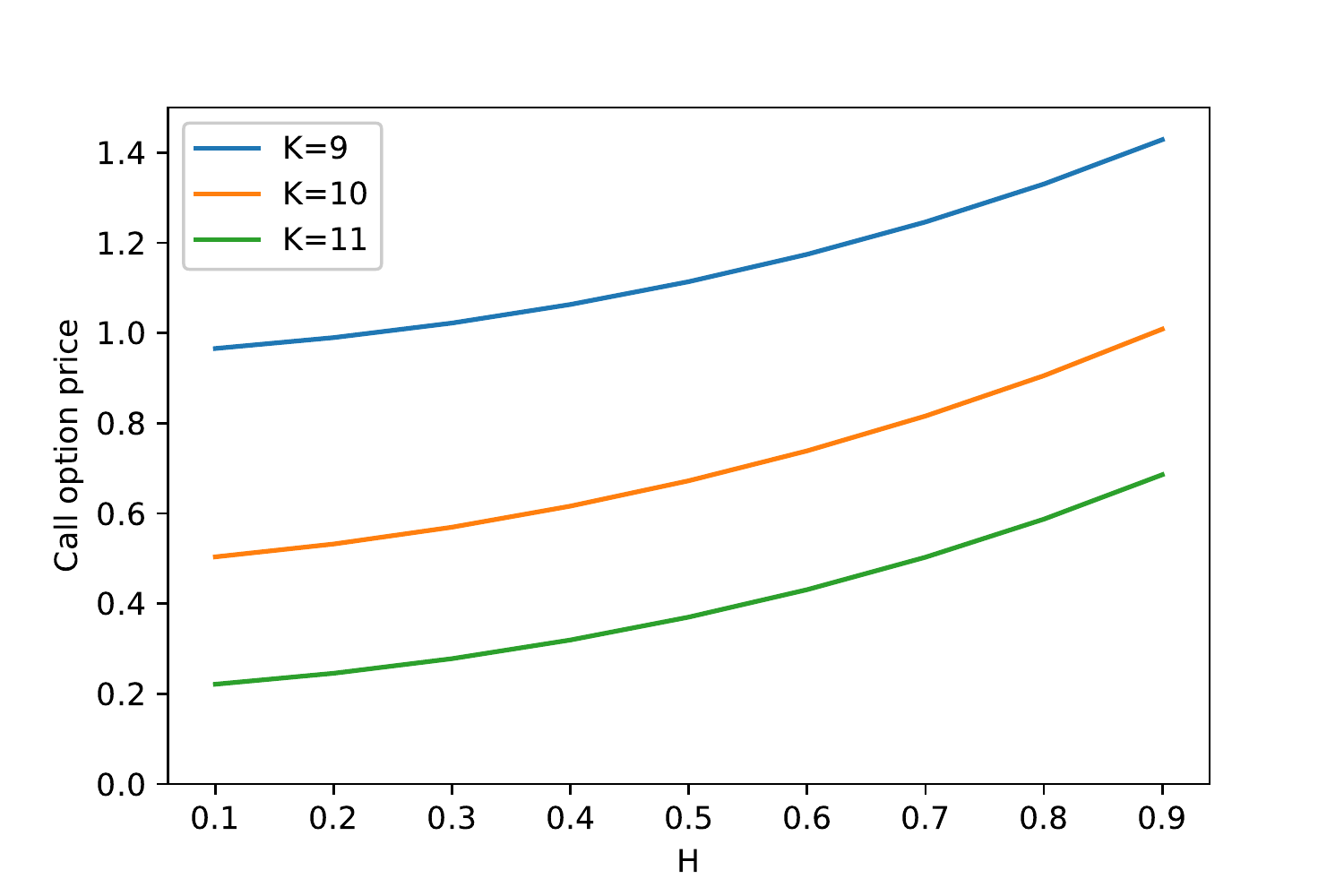}
	\caption{Call option}
	\end{subfigure}
	\begin{subfigure}{0.44\textwidth}
	\includegraphics[width=\textwidth]{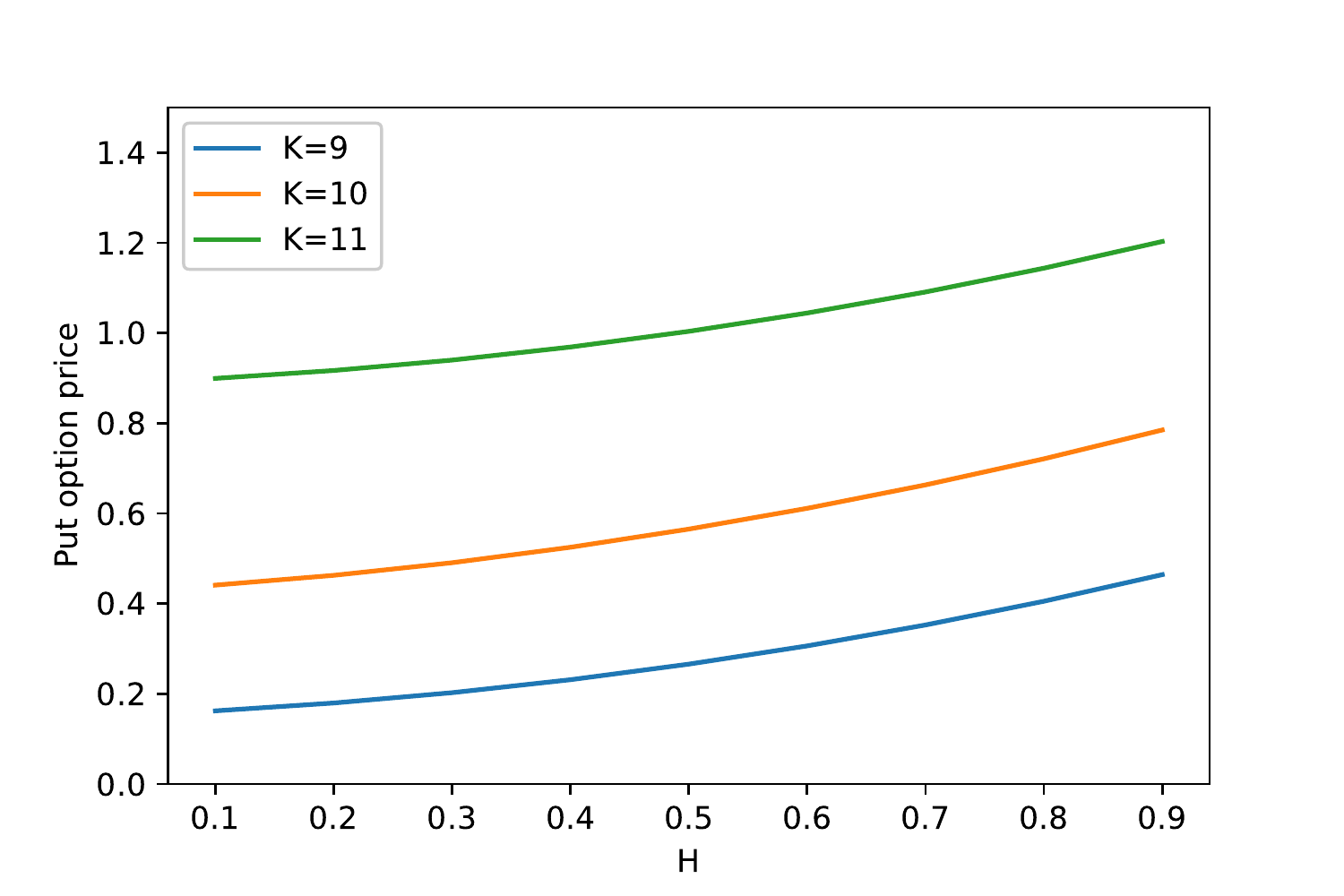}
	\caption{Put option}
	\end{subfigure}
	\caption{European option values under the polynomial process with different $H$ and $K$, $t=0$ and $T=3$ fixed.}
	\label{fig.price_poly}
\end{figure}

We further present the prices of European call and put options under the GfOU process in \Cref{fig.price_GfOU} and the polynomial process in \Cref{fig.price_poly}, respectively, for different Hurst indices $H$ and strike prices $K$. Obviously, the Hurst index $H$ has a significant impact on the option price, which indicates the importance of modeling the asset price as a stochastic process with a suitable $H$.

\begin{table}[!htb]
  \centering
  \caption{Errors of the COS method for a European call option under the GfOU process with $K=10$, $t=0$, $T=3$, $r=0.1$. }
  \resizebox{0.98\textwidth}{!}{
    \begin{tabular}{cccccccccc}
    \toprule
    \diagbox {$L$}{$H$} & 0.1 & 0.2 & 0.3 & 0.4 & 0.5 & 0.6 & 0.7 & 0.8 & 0.9 \\
    \midrule
    4   & 8.4e-02 & 9.0e-02 & 9.8e-02 & 1.1e-01 & 1.2e-01 & 1.3e-01 & 1.5e-01 & 1.6e-01 & 1.8e-01 \\
    8   & 1.8e-03 & 2.0e-03 & 2.1e-03 & 2.4e-03 & 2.6e-03 & 2.9e-03 & 3.3e-03 & 3.7e-03 & 4.1e-03 \\
    16  & 2.8e-08 & 2.5e-08 & 2.1e-08 & 1.6e-08 & 9.1e-09 & 1.0e-09 & 8.6e-09 & 2.0e-08 & 3.2e-08 \\
    \bottomrule
    
    \end{tabular}}
  \label{table.cos_error}
\end{table}

\section{Conclusion}
\label{sec.6}
In this paper, we have shown that, on the basis of several transformations of variables and the analytic continuation technique, integrals that define the conditional expectation, variance and characteristic function of stochastic processes, with respect to fractional Brownian motion (fBm), can be reformulated so that they exist for all relevant Hurst indices.  
Numerical experiments for the conditional variance of the fOU process, and to recover the probability density function of certain stochastic processes that can be derived from the fOU process, confirm the robustness and efficiency of the integral formulations and of the numerical technique, regarding different values of the Hurst index. Moreover, by means of the COS method, we have presented  accurate and highly efficient option pricing results for processes  connected to the fOU process, on the basis of the conditional characteristic function of the fOU process.

The methodology presented here can be used as an alternative to Monte Carlo simulation when appropriate, or as a validation technique for implementations of other numerical techniques. 
\section*{Acknowledgments}	
F. Gao would like to thank the China Scholarship Council (CSC, No. 202006280439) for the financial support. N. M. Temme acknowledges financial support from {\emph{Ministerio de Ciencia e
Innovaci\'on}}, Spain,  project MTM2012-11686; N. M. Temme thanks CWI, Amsterdam, for general support.
\section*{Acknowledgments}	
F. Gao would like to thank the China Scholarship Council (CSC, No. 202006280439) for the financial support. N. M. Temme acknowledges financial support from {\emph{Ministerio de Ciencia e
Innovaci\'on}}, Spain,  project MTM2012-11686; N. M. Temme thanks CWI, Amsterdam, for general support.

\begin{appendices}
\section{Analytic continuation}
\label{appendix.analytic continuation}
In complex analysis, analytic continuation is a classical technique to extend the domain of definition of a given analytic function. Analytic continuation often succeeds in defining further values of a function, for example, in a new region where an infinite series representation, in terms of which it is initially defined, becomes divergent.

As a standard example to explain analytic continuation for a class of integrals, we consider the function $F(x)$ defined by
\[
F(x)=\frac{1}{\Gamma(x)}\int_0^\infty t^{x-1} e^{-t} f(t)\,\d t,
\]
where the function $f(t)$ is analytic in $[0,\infty)$, and in a complex domain around this interval.  $F(x)$ is an analytic function in the domain $\Re \{x\}>0$. Next, we assume that $f(0)\ne0$ and, in that case, the integral does not converge if $x=0$. The reciprocal gamma function in front of the integral will allow us to give an  interpretation to $F(x)$  for $x\le0$. 

We first take $f(t)=1$ for $t\ge0$. In that case, $F(x)=1$, and we see that the combination of the front factor and the integral, considered as one quantity, becomes a function which is defined for all real and complex values of $x$. A second step is writing $f(t)=f(0)+\left(f(t)-f(0)\right)$, or introducing the functions
\[
g(t)=\frac{f(t)-f(0)}{t},\quad g(0)=f^\prime(0), \quad G(x)=f(0)+\frac{1}{\Gamma(x)}\int_0^\infty t^{x} e^{-t} g(t)\,\d t.
\]
The function $G(x)$ is an analytic function  for $\Re \{x \}>-1$, and $F(x)=G(x)$ for $\Re \{x\}>0$, and we conclude that $F(x)$ which was initially defined for $\Re \{x\}>0$, has $G(x)$ as its analytic continuation for $\Re \{x\}>-1$, and $F(x)=G(x)$ for $\Re \{x\}>-1$.

For further details on analytic continuation, we refer to the references \cite[Section~1.10(ii)]{Roy:2010:AAM} and  \cite[Chapter~IV]{Titchmarsh:1958:TTF}.

A second method is based on integration by parts. We have, using $x\Gamma(x)=\Gamma(x+1)$,  for $\Re (x)>0$,
\[
F(x)=\frac{1}{\Gamma(x+1)}\int_0^\infty e^{-t} f(t)\,\d\left(t^x\right)=\frac{1}{\Gamma(x+1)}\int_0^\infty t^x e^{-t}\left(f(t)-f^\prime(t)\right)\,\d t.
\]
The new integral is defined for $\Re \{x\}>-1$, and, again, we have obtained the analytic continuation of $F(x)$ from $\Re \{x\}>0$ to $\Re \{x\}>-1$.

We can repeat these two methods to obtain for $F(x)$ the analytic continuation for $\Re \{x\}>-2$, $\Re \{x\}>-3$, and so on.

\section{The case \texorpdfstring{$s=0$}{s=0}}
\label{appendix.s0}
In \eqref{Hz}, we see that we do not need to handle a singularity at the endpoint $z=s$ when $s>0$. However,  this will be needed when $s=0$, in which case the variable of integration, $z$  in \eqref{hz}, will take the value $z=0$.

To deal with this, first of all, we write the representation of $R_n(\kappa,z)$ in \eqref{eq:G09} in a different form, because the ${}_2F_1$-function is not defined at $z=0$ (although the front factor $z^{2\kappa+n}$ will control the product of the two quantities).

We use 
\begin{equation}\label{eq:sz0:01}
w_1(z)=A w_3(z)+Bw_4(z)
 \end{equation}
of \cite[Eqn.~15.10.21]{Olde:2010:GHF}, with $w_3(z)$ and $w_4(z)$ in the first lines of Equations (15.10.13) and (15.10.14) of that reference. Here,
\begin{equation}\label{eq:sz0:02}
A=\frac{\Gamma(\kappa+1)\Gamma(-n-2\kappa)}{\Gamma(-n-\kappa)}=\frac{\Gamma(\kappa+1)\Gamma(n+1+\kappa)}{2\cos (\pi\kappa)\Gamma(n+1+2\kappa)},
\end{equation}
and
\begin{equation}\label{eq:sz0:03}
B=\frac{\Gamma(\kappa+1)\Gamma(n+2\kappa)}{\Gamma(n+2\kappa+1)\Gamma(\kappa)}=\frac{\kappa}{n+2\kappa}.
\end{equation}
For $w_3(z)$, we have an elementary form, i.e.,
\begin{equation}\label{eq:sz0:04}
w_3(z)=\FG{n+2\kappa+1}{\kappa}{n+2\kappa+1}{\frac{z}{t}}=(1-z/t)^{-\kappa},
\end{equation}
and $w_4(z)$ is given by
\begin{equation}\label{eq:sz0:05}
w_4(z)=\left(\frac{z}{t}\right)^{-n-2\kappa}\FG{-\kappa-n}{1}{1-n-2\kappa}{\frac{z}{t}}.
\end{equation}
This gives us, for \eqref{eq:G09},
\begin{equation}\label{eq:sz0:06}
R_n(\kappa,z)=\frac{\Gamma(\kappa+1)\Gamma(n+1+\kappa)}{2\cos (\pi\kappa)\Gamma(n+1+2\kappa)}z^{2\kappa+n}
+\frac{\kappa (t-z)^{\kappa} t^{n+\kappa}}{n+2\kappa}\FG{-\kappa-n}{1}{1-n-2\kappa}{\frac{z}{t}}.
\end{equation}

This ${}_2F_1$-function is not defined for  $z\to t$ when $\kappa>0$. In that case \Cref{eq:G09} should be used.

\end{appendices}

\bibliographystyle{plain}
\bibliography{ref}
\end{document}